\documentclass{IEEEtran}
\usepackage{cite}
\usepackage{amsmath,amssymb,amsfonts}
\usepackage{graphicx}
\usepackage{textcomp,nicefrac}

\usepackage{threeparttable}
\usepackage{lineno}
\usepackage{xcolor}
\usepackage{hyperref}
\hypersetup{
    colorlinks=true,
    linkcolor=blue,
    citecolor=blue,
    urlcolor=cyan!40!blue
}
\def\BibTeX{{\rm B\kern-.05em{\sc i\kern-.025em b}\kern-.08em
T\kern-.1667em\lower.7ex\hbox{E}\kern-.125emX}}
\begin{document}
\title{Study of the intrinsic resolution of LaBr$_3$(Ce,Sr) and NaI(Tl) crystals}
\author{Peiyi Feng, Xilei Sun, Zhenghua An, Dali Zhang, Xinqiao Li, Shaolin Xiong, and Hong Lu
\thanks{This work was supported by the National Key Research and Development Program of China (Nos. 2022YFB3503600 and 2021YFA0718500), Strategic Priority Research Program of the Chinese Academy of Sciences (No. XDA15360102), and National Natural Science Foundation of China (Nos. 12173038%李新乔
, 12273042%熊少林
, and 12075258%卢红
). (Corresponding author: Peiyi Feng, Xilei Sun, and Zhenghua An.)}
\thanks{Peiyi Feng is with the Institute of High Energy Physics, Chinese Academy of Sciences, and also with the University of Chinese Academy of Sciences, both located at Beijing 100049, China (e-mail: fengpeiyi@ihep.ac.cn).}
\thanks{Xilei Sun, Zhenghua An, Dali Zhang, Xinqiao Li, Shaolin Xiong, and Hong Lu are with the Institute of High Energy Physics, Chinese Academy of Sciences, Beijing 100049, China.}}

% (e-mail: sunxl@ihep.ac.cn, anzh@ihep.ac.cn, zhangdl@ihep.ac.cn, lixq@ihep.ac.cn, xiongsl@ihep.ac.cn, luh@ihep.ac.cn)

% non-proportionality

\maketitle

\begin{abstract}

The Gravitational wave burst high-energy Electromagnetic Counterpart All-sky Monitor (GECAM) utilizes a large number of LaBr$_3$ and NaI(Tl) crystals as sensitive materials for its gamma-ray detectors. To address the fitting issues of the energy resolution curves in the ground calibration of the GECAM detectors, this work conducts a comprehensive testing and comparative study of the energy resolution of 1-inch LaBr$_3$(Ce,Sr) and NaI(Tl) crystals produced from the same batch. We employed a Hard X-ray Calibration Facility (HXCF), a PMT single-photoelectron calibration system, and Geant4 Monte Carlo simulation tools to quantify seven factors influencing energy resolution. The results indicate that the contributions of various components to energy resolution differ, with photoelectron statistical fluctuations and intrinsic resolution being predominant. For 100 keV X-rays, the total energy resolution of the LaBr$_3$(Ce,Sr) crystal is 3.71\% ± 0.03\% (expressed as 1-$\sigma$), with a contribution from photoelectron statistical fluctuations of 2.69\% ± 0.00\% and an intrinsic resolution of 2.45\% ± 0.05\%. For 100 keV X-rays, the total energy resolution of the NaI(Tl) crystal is 4.41\% ± 0.14\% (expressed as 1-$\sigma$), with a contribution from photoelectron statistical fluctuations of 3.20\% ± 0.00\% and an intrinsic resolution of 2.90\% ± 0.21\%. We discussed the sources of intrinsic resolution, and the results indicate that the intrinsic resolution of the LaBr$_3$(Ce,Sr) crystal primarily arises from luminescence non-proportionality, while that of the NaI(Tl) crystal mainly stems from fluctuations during energy transfer. This study emphasizes precise and specific experimental measurements and comparative research, demonstrating that both factors are important contributors to intrinsic resolution.
% 引力波暴高能电磁对应体全天监测器（GECAM）使用了大批量的LaBr3和NaI晶体作为其伽马射线探测器的灵敏材料。为了解决GECAM探测器地面标定中能量分辨率曲线的拟合问题，本工作对同一批次生产的1英寸LaBr3(Ce,Sr)和NaI(Tl)晶体的能量分辨率进行了全面的测试和对比研究。我们使用硬X射线地面标定装置（HXCF）、PMT单光电子刻度系统以及Geant4蒙特卡洛模拟工具，量化了影响能量分辨率的七种因素。结果表明，能量分辨率中各种成分占比不一，以光电子统计涨落的贡献和本征分辨率为主。对于100keV X射线，LaBr3(Ce,Sr)晶体的总能量分辨率为3.71%+-0.03%，光电子统计涨落的贡献为2.95%+-0.01%，而本征分辨率是2.10%+-0.06%。对于100keV X射线，NaI(Tl)晶体的总能量分辨率为4.41%+-0.14%，光电子统计涨落的贡献为3.20%+-0.01%，而本征分辨率是2.90%+-0.21%。
% 我们讨论了本征分辨率的来源问题，结果表明LaBr3(Ce,Sr)晶体的本征分辨率主要来源于发光非线性，而NaI(Tl)晶体的本征分辨率主要来源于能量传递过程的涨落。本研究侧重精确而具体的实验测量和对比研究，表明这二者皆为本征分辨率的重要来源。

\end{abstract}


% \linenumbers
% \setlength\linenumbersep{0.05cm}

\section{Introduction}
\label{sec:introduction}

% \IEEEPARstart{N}{aI(Tl)} crystal is a well-established inorganic scintillator known for its excellent luminescent properties. It is easy to process into various shapes and sizes and has a low manufacturing cost \cite{hawrami2022growth, wang2023segregation}. The discovery of NaI(Tl) crystals marked a groundbreaking development in high-energy photon detection technology, and their application in high-energy physics has a history spanning several decades, making the technology highly mature \cite{suerfu2020growth, kim2019limits, mao2007optical, yang2015improving, wakabayashi2015applicability}. LaBr$_3$ crystals, a novel type of inorganic scintillator, exhibit superior scintillation characteristics, such as high energy resolution, high light output, and fast decay times \cite{kumar2009efficiency, mazumdar2013studying, dhibar2018characterization}. LaBr$_3$ is considered an ideal replacement for traditional NaI(Tl) crystals and is widely used in radiation detection fields, including high-energy physics.
% NaI晶体是一种性能优良的传统无机闪烁晶体，具有出色的发光性能，且易于加工成多种形状和尺寸，制造成本低。NaI晶体的发现是高能光子探测技术中划时代的突破，在高能物理中的应用已经有几十年的历史，技术方面相当成熟。LaBr3晶体是一种新型的无机闪烁晶体，它表现出优异的闪烁性能，如高能量分辨率、高光输出和快速衰减时间。LaBr3晶体是传统闪烁晶体NaI的理想替代品，被广泛用于高能物理等辐射探测领域。

\IEEEPARstart{T}{he} Gravitational wave burst high-energy Electromagnetic Counterpart All-sky Monitor (GECAM) was developed specifically to monitor high-energy electromagnetic events such as gamma-ray bursts  (GRBs), fast radio bursts (FRBs), and magnetar flares \cite{SSPMA-2019-0417, SSPMA-2020-0457, zhang2022dedicated, zhang2019energy, li2021technology, zhang2023performance, wang2024simulation}. Since its launch, GECAM has achieved a series of original results in various fields related to gamma-ray transients \cite{zhao2023gecam, zhao2023paired}. GECAM employs LaBr$_3$ and NaI(Tl) crystals as the sensitive materials for its gamma-ray detectors (GRDs). The observation and analysis of the brightest gamma-ray burst to date, GRB 221009A, and the second-brightest, GRB 230307A, represent some of GECAM's most significant achievements \cite{zheng2024observation, an2023insight, sun2023magnetar}. These important and intriguing discoveries demonstrate the outstanding performance of LaBr$_3$ and NaI(Tl) crystals in the field of high-energy astrophysics. The successful application of LaBr$_3$ crystals in GECAM marks the first large-scale use of this novel scintillator in astrophysics, paving the way for its broader application in other fields.
% 引力波暴高能电磁对应体全天监视器（GECAM）被研发专用于监测伽马射线暴、快速射电暴和磁星耀发等高能电磁事件。自GECAM发射以来，它在伽马射线暂现源的各个领域取得了一系列原创性成果。GECAM采用LaBr3和NaI(Tl)晶体作为其伽马射线探测器（GRD）的灵敏材料。著名的迄今为止最亮的伽马射线暴GRB 221009A和第二亮的伽马射线暴GRB 230307A的观测和分析是GECAM最具代表性的成果，这些重要而有趣的原创性发现展示了LaBr3和NaI晶体在高能天体物理领域极为出色的表现。LaBr3晶体耦合硅光电倍增管（SiPM）这一探头设计方案在GECAM卫星上的成功应用也是该新型闪烁晶体首次在天文物理方面的大规模应用，将推动其在更广泛的领域发挥重要作用。
% coupled with silicon photomultiplier (SiPM) arrays

To investigate the non-proportionality issues in LaBr$_3$ and NaI(Tl) crystals during low-energy gamma-ray detection and to address the E-C (energy-channel) relationship errors in the calibration of GECAM satellite detectors, we have conducted a detailed study of the energy response of these crystals \cite{FPY}. In the ground calibration of the Gamma-ray Transient Monitor (GTM, also known as GECAM-D, the fourth member of the GECAM series), the NaI(Tl) crystal exhibited poor energy resolution in the medium- to high-energy range, which limited the calibration energy range \cite{feng2024detector}. To better fit the energy resolution curve, it is critical to investigate the factors affecting the energy resolution of NaI(Tl) crystals. As a follow-up to previously published work \cite{FPY}, this study specifically compares the energy resolution of 1-inch LaBr$_3$(Ce,Sr) and NaI(Tl) crystal samples, produced in the same batch as the GRDs, and presents their ultimate resolution—intrinsic resolution.

Energy resolution is one of the key parameters characterizing the performance of scintillators. There are seven potential sources of total energy resolution $\sigma/E$ (expressed as 1-$\sigma$): (1) fluctuations in the energy transfer process $\delta_{trans}$, (2) non-proportional luminescence $\delta_{non}$, (3) uneven photon collection $\delta_{un}$, (4) contributions from photoelectron statistical fluctuations $\delta_{st}$, (5) single-photoelectron resolution of the photomultiplier tube (PMT) $\delta_{spe}$, (6) electronic noise $\delta_{noise}$, and (7)  temperature drift $\delta_{temp}$ \cite{deng2022exploring, bissaldi2009ground}. Their relationship is described by Equation~\ref{eq:1}. Since sources (1) and (2) are determined by the inherent properties of the crystal material, they are collectively referred to as the intrinsic resolution $\delta_{int}$, which is calculated by Equation~\ref{eq:2}. 
% 能量分辨率是表征闪烁体性能的重要参数之一。能量分辨率有七种可能的来源：（1）能量传递过程的涨落、（2）非线性的发光、（3）光收集不均匀性、（4）光电子统计涨落的贡献、（5）光电倍增管的单光电子分辨率、（6）电子学噪声、（7）温度漂移。它们的关系如公式1。由于（1）（2）由材料本身的性质决定，我们将二者称为本征分辨率，本征分辨率由公式2计算。

\begin{equation}\label{eq:1}
(\sigma/E)^2 = \delta_{int}^2 + \delta_{un}^2 + \delta_{st}^2 + \delta_{spe}^2 + \delta_{noise}^2 + \delta_{temp}^2.
\end{equation}

\begin{equation}\label{eq:2}
\delta_{int}^2 = \delta_{trans}^2 + \delta_{non}^2 .
\end{equation}

Intrinsic resolution is the important component of the total energy resolution \cite{moszynski2004intrinsic, moszynski2016energy}. S.A. Payne et al. theoretically studied the intrinsic energy resolution and non-proportionality of scintillator detectors \cite{payne2011nonproportionality, payne2015nonproportionality}, while V. Ranga and P. Limkitjaroenporn used the Wide-Angle Compton Coincidence (WACC) technique to measure the light yield non-proportionality and intrinsic energy resolution of several inorganic scintillators \cite{kaintura2021energy, 2017Intrinsic, LIMKITJAROENPORN201815110, 2009A, 2012Non}. Y. Deng et al. demonstrated that the non-proportionality of liquid scintillators has only a weak contribution to the intrinsic resolution for electrons \cite{deng2022exploring}, with most of the intrinsic resolution likely originating from fluctuations in the energy transfer process. However, studies have shown that the intrinsic resolution of LaBr$_3$(Ce) crystals primarily arises from their non-proportional energy response \cite{swiderski2010energy, sriwongsa2019non}. This indicates that there is a distinction between crystal scintillators and liquid scintillators. However, does the main source and contribution proportion of intrinsic resolution differ for different types of crystal scintillators? 
% 本征分辨率是闪烁体总能量分辨率的重要成分，也是限制总能量分辨率的主要原因。S.A. Payne等人从理论上研究了闪烁体探测器的本征能量分辨率和非线性。V. Ranga和P. Limkitjaroenporn等人使用广角康普顿符合（WACC）技术测量了一些无机闪烁体的光产额非线性和本征能量分辨率。邓勇等人阐述了液体闪烁体的非线性对电子的本征分辨率只有微弱的贡献，大部分本征分辨率可能来自于能量转移过程中的涨落。然而，研究表明了LaBr3（Ce）晶体的本征分辨率主要来自非线性的能量响应。这说明晶体与液闪存在不一样的情况。然而，对于不同类型的晶体，它们的本征分辨率的主要来源与成分占比是否也存在差异呢？

In fact, precise experimental measurements of the intrinsic resolution of crystals are scarce in the published paper. Most researchers have only considered the impact of statistical fluctuations in photoelectrons while neglecting other factors \cite{2017Intrinsic, kaintura2021energy}. This simplified approach leads to an overestimation of the calculated intrinsic resolution. We previously conducted a preliminary study on the intrinsic resolution of LaBr$_3$(Ce) \cite{feng2023intrinsic}, while this work focuses on a detailed investigation of the intrinsic resolution of LaBr$_3$(Ce,Sr) and NaI(Tl) crystals. We utilized two hard X-ray ground calibration facilities (HXCFs) and single-photoelectron calibration methods, and simulated the energy deposition and photon transport processes in LaBr$_3$(Ce,Sr) and NaI(Tl) crystals using Geant4 software, quantifying the effects of the seven aforementioned factors on the total energy resolution. We investigated the relationship between the energy transfer process, luminescence non-proportionality, and intrinsic resolution, aiming to clarify the concept of intrinsic resolution, determine its magnitude, and discuss its sources, with the goal of further identifying the limiting factors of energy resolution.
% 事实上，在已发表的文章中，对于晶体本征分辨率的精确实验测量非常少。多数研究者只考虑了光电子统计涨落的影响，而忽略了其他的因素。这种简单的处理使得计算出来的本征分辨率偏大。我们在之前已经简单研究了LaBr$_3$(Ce)本征分辨率，而本工作专门详细研究LaBr$_3$(Ce,Sr) and NaI(Tl) 晶体的本征分辨率。我们使用硬X射线地面标定装置(HXCF)、单光电子刻度方法，并使用Geant4软件模拟了入射粒子在NaI(Tl)晶体中能量沉积和发光的过程，量化了上述七种因素对总能量分辨率的贡献。我们研究能量传递过程、发光非线性与本征分辨率的关系，旨在阐明本征分辨率的概念、确定本征分辨率的大小，并讨论本征分辨率的来源问题，进一步明确能量分辨率的限制因素。

\section{Experimental Setups}

This section introduces the experimental setups used in this study, including two hard X-ray ground calibration facilities (HXCFs) and a PMT single-photoelectron calibration (SPEC) system. The HXCFs are primarily used to test the energy response of the LaBr$_3$(Ce,Sr) and NaI(Tl) crystals to X-rays, including energy spectra, energy-channel (E-C) relationship, and energy resolution. Furthermore, the single-crystal HXCF is employed to assess the radial ($X$- and $Y$-axes) non-uniformity of the crystals. The SPEC system is used to calibrate the single-photoelectron response of the PMT coupled with the crystals, enabling the calculation of the crystals' absolute light yield.
% 本章节介绍了本研究所使用的实验装置，包括两套硬X射线地面标定装置（HXCF）和一套PMT单光电子刻度（SPEC）装置。HXCF主要用于测试LaBr$_3$(Ce,Sr)和NaI(Tl)晶体对X射线的能量响应，包括能谱、E-C关系、能量分辨率。此外，HXCF还用于测试NaI(Tl)晶体的不均匀性。SPEC装置则用于刻度与晶体耦合的PMT的单光电子响应，以计算晶体的绝对光产额。

\subsection{Hard X-ray Ground Calibration Facility}
\label{chap:Hard X-ray Ground Calibration Facility}

\begin{figure*}[!htb]
\centering
\includegraphics
  [width=0.75\hsize]
  {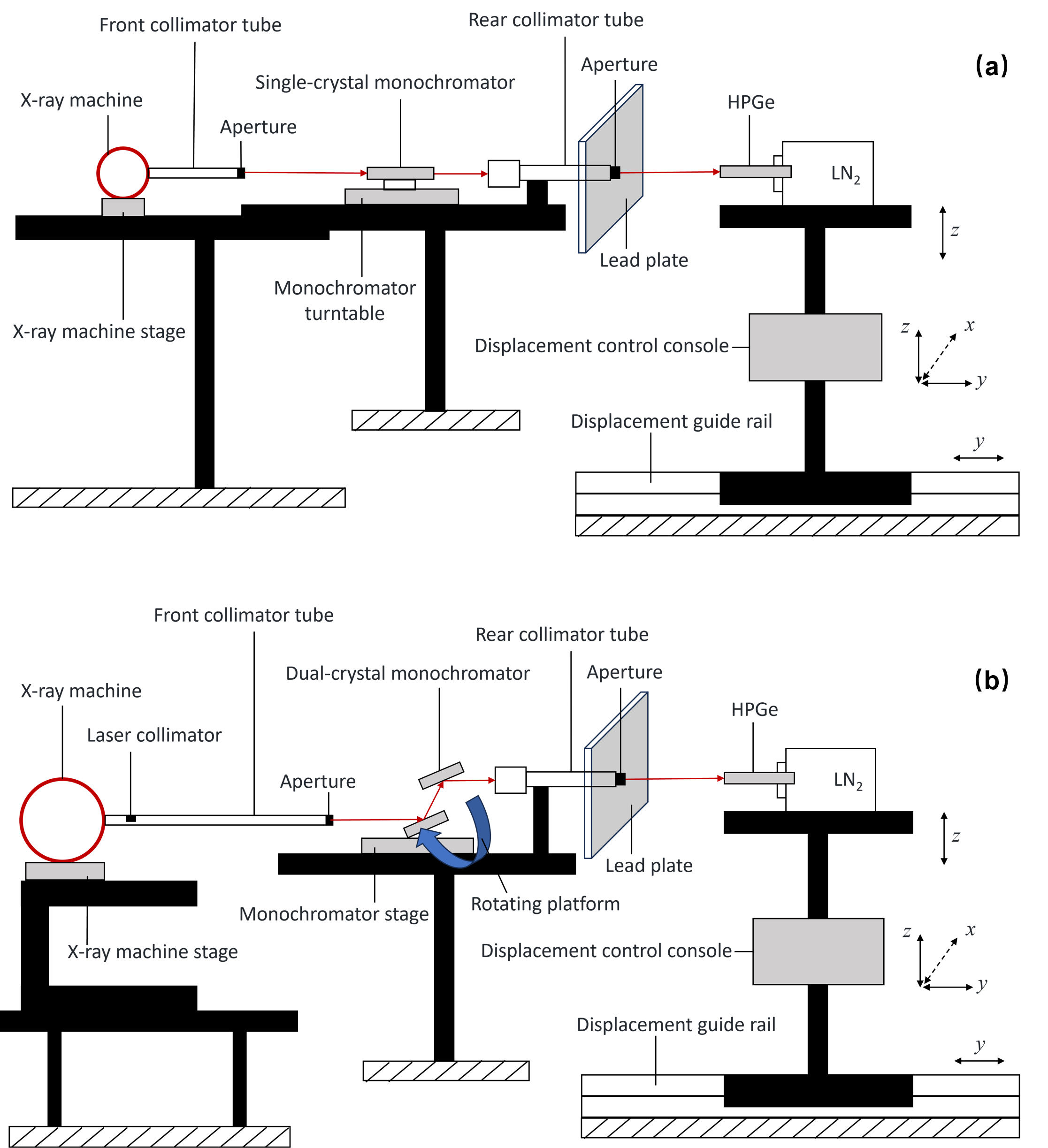}
\caption{(a) A hard X-ray ground calibration facility (HXCF) based on a single-crystal testing system. (b) A hard X-ray ground calibration facility (HXCF) based on a dual-crystal testing system.}
\label{fig:1}
\end{figure*}

\begin{figure*}[!htb]
\centering
\includegraphics
  [width=0.75\hsize]
  {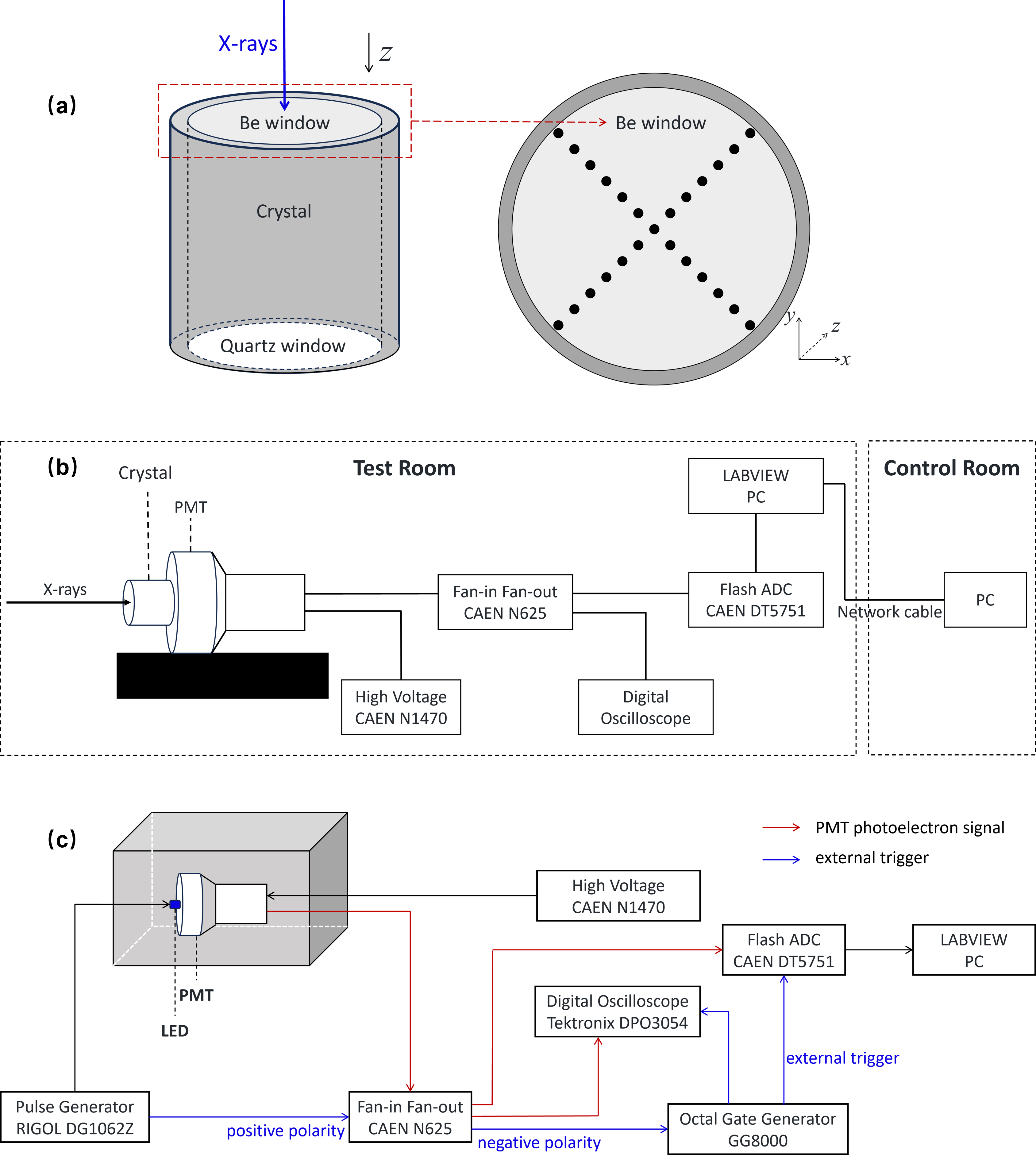}
\caption{(a) Left: a 3D schematic diagram of the 1-inch LaBr$_3$(Ce,Sr) and NaI(Tl) crystal structure. Right: 25 position points distributed on the beryllium window surface of the crystals, used to test their uniformity. (b) Data acquisition system for the LaBr$_3$(Ce,Sr) and NaI(Tl) crystal detector. (c) A system for PMT single-photoelectron calibration based on pulse voltage-driven LED illumination, where red and blue markers represent the photoelectron signal from the PMT and the external trigger signal of the data acquisition device, respectively.}
\label{fig:2}
\end{figure*}

We used two HXCFs to test the energy response of the LaBr$_3$(Ce,Sr) and NaI(Tl) crystals to X-rays in the 10-100 keV energy range. These HXCFs, located at the National Institute of Metrology (NIM) in Changping, Beijing, were originally designed for the high-energy Hard X-ray Modulation Telescope (HXMT) project and have played a crucial role in the ground calibration of gamma-ray detectors for the GECAM series satellites and the Space-based multi-band astronomical Variable Objects Monitor (SVOM) satellite \cite{feng2024detector, 2021Ground, 2021Calibration, 2019Ground, zhang2020overview}. The two HXCF setups are the single-crystal and dual-crystal monochromatic X-ray test systems.
%  \cite{2021The, 2019The, 2022Research}
% 我们使用了两套硬X射线地面标定装置(HXCFs)测试了LaBr$_3$(Ce,Sr)和NaI(Tl)晶体对10-100keV X射线的能量响应。HXCF位于中国北京昌平的国家计量院（NIM），该测试系统最初为高能硬X射线调制望远镜（HXMT）建立，在GECAM系列卫星和空间可变天体监视器（SVOM）卫星的伽马射线探测器地面标定中也发挥了重要作用。两套HXCFs分别是单能X射线单晶测试装置和单能X射线双晶测试装置。

The single-crystal testing system (Fig.~\ref{fig:1} (a)) consists of an X-ray generator, a single-crystal monochromator, lead collimators, and a standard detector. The X-ray generator uses an Oxford X-ray tube with a copper anode target, a maximum tube voltage of 50 kV, and a maximum tube current of 1 mA. The single-crystal monochromator includes a LiF crystal, a crystal fixation structure, and a rotating stage. The lead collimator is composed of a front and rear collimator, which also serves to block stray light from the X-ray generator. The standard detector is a low-energy HPGe detector manufactured by Canberra, used to determine the flux and energy of the monochromatic X-ray beam. This HPGe detector has been calibrated for energy linearity, energy resolution, and detection efficiency using standard radioactive sources such as $^{55}$Fe, $^{133}$Ba, $^{241}$Am, $^{57}$Co, $^{109}$Cd, and $^{152}$Eu, with an energy detection range of 5-300 keV \cite{2016LEGe}.
% 单晶测试装置由X射线光机、单晶单色仪、铅准直器和标准探测器组成。X射线光管采用牛津光管，阳极靶为铜靶，最大管电压为50 kV，最大管电流为1 mA。单晶单色仪包括一块LiF晶体、晶体固定结构和旋转转台。铅准直器由前准直器和后准直器构成，兼具屏蔽X射线光机出口杂散光的功能。标准探测器采用堪培拉公司生产的低能HPGe探测器，用于确定单能X射线束流的通量和能量。该HPGe探测器已使用标准放射源55Fe、133Ba、241Am、57Co、109Cd和152Eu对其能量线性、能量分辨率及探测效率进行标定，探测能区为5-300 keV。

The dual-crystal testing system (Fig.~\ref{fig:1} (b)) consists of an X-ray generator, a dual-crystal monochromator, lead collimators, and a standard detector. The X-ray generator is the Y.TU225-D02 model manufactured by YXLON, featuring a tungsten anode target, a maximum tube voltage of 225 kV, and a maximum power output of 3000 W. The dual-crystal monochromator comprises two Si crystals, a T-shaped structure, and a high-precision angle meter. In addition to the front and rear collimators, the X-ray generator is equipped with a laser collimator to simulate the beam path of the X-rays. The standard detector for the dual-crystal testing system is identical to that of the single-crystal system, utilizing an HPGe detector. Both testing systems operate on the same principle, producing monochromatic X-rays through Bragg diffraction. The key difference is that the single-crystal system offers higher flux and better monochromaticity at lower energy ranges, with the proportion of monochromatic light exceeding 97\% in the 5-30 keV energy range.
% 双晶测试装置由X射线光机、双晶单色仪、铅准直器和标准探测器组成。X射线光机采用YXLON公司生产的Y.TU225-D02型号，其X射线光管阳极靶为钨靶，最大管电压为225 kV，最大功率为3000 W。双晶单色仪由两块Si晶体、T型结构和高精度角度仪组成。准直系统除了前准直器和后准直器外，X射线光机出口还配备了激光准直器，用于模拟X射线束流的光路。双晶测试装置的标准探测器与单晶测试系统相同，均采用HPGe探测器。两套测试装置的工作原理相同，均通过布拉格衍射产生单能X射线。不同之处在于，单晶测试系统在低能区通量更大且单色性更好，在5-30 keV能区，单色光比例超过97%。

We used the single-crystal testing system to measure the X-ray response in the 10-40 keV energy range, and the dual-crystal testing system for the 40-100 keV range. The crystal to be tested was placed on a displacement platform, aligned horizontally with the HPGe detector. A lead plate was positioned between the detector and the X-ray source to shield against stray radiation from the X-ray generator, with a 2 mm diameter aperture at the center of the lead plate. Before each test, the monochromator angles were adjusted to generate monochromatic X-rays of different energies using the Bragg diffraction principle. We recorded the HPGe detector spectral data collected by the multi-channel analyzer using GENIE2000 spectrum analysis software, with a collection time of 100 seconds. By analyzing the data from the HPGe detector, we determined the energy and flux of the X-ray beam. With the monochromator angle, tube voltage, and current held constant, the platform was adjusted to direct the monochromatic X-rays onto the crystal under test. Since this study does not involve efficiency issues, the acquisition time of the crystal detectors under test is not considered. It is only necessary to ensure that the count exceeds 20,000, and at least 10 sets of background data are collected during the experiment.
% 我们使用了单晶测试装置来测量10-40keV能区的X射线响应，并使用了双晶测试装置来测量40-100keV的X射线响应。待测试的晶体被放置在HPGe探测器所在的位移平台上，并与HPGe探测器保持在同一水平线上。在探测器与X射线出口之间放置了一块铅板，用于屏蔽从X光机出来的杂散光，铅板中心设置了一个直径为2mm的小孔径光阑口。在每次测试前，需要调节单色仪的不同角度，利用布拉格衍射原理获得不同能量的单能X射线。我们使用GENIE2000能谱分析软件记录从多道分析器中采集到的HPGe探测器谱数据，采集时间为100秒。通过分析HPGe探测器的数据，确定了X射线束流的能量和通量。保持单色仪角度、管电压和管电流不变，移动平台使单能X射线入射到待测晶体上。由于本研究不涉及效率问题，因此我们不考虑被测试的晶体探测器的采集时间，只需保证计数大于20000个，且在实验过程中测试不少于10组本底数据。

This study used pre-encapsulated 1-inch cylindrical LaBr$_3$(Ce,Sr) and NaI(Tl) crystal samples. The incident window was made of a 0.2 mm-thick beryllium sheet, and the crystal was surrounded by a 1.5 mm-thick aluminum alloy casing, with a light output window equipped with 3 mm thick-quartz glass. The incident and cylindrical surfaces were coated with 0.4 mm-thick PTFE (Teflon) reflective material to enhance the collection efficiency of scintillation photons (Fig.~\ref{fig:2} (a)). The crystals under test were coupled to a Hamamatsu CR160 PMT using silicone grease. During the tests, the PMT was operated at -800 V for LaBr$_3$(Ce,Sr) and -1000 V for NaI(Tl) crystal. Figure.~\ref{fig:2} (b) shows a block diagram of the data acquisition system for the tested crystal detectors. X-ray beams entered through the beryllium window, interacted with the crystals, and generated scintillation photons, which were collected at the PMT photocathode and converted into photoelectrons  \cite{han2023csi, liu2023toward, zhang2022transition}. After being multiplied by the dynodes, a charge pulse signal was produced. We used a Flash ADC DT5751 digitizer to record the signal with LABVIEW software on a laboratory computer \cite{LI2014Particle}, which was connected via an Ethernet cable to a control room computer for data acquisition control. The ambient temperature was maintained at 20±1°C. For the 10-100 keV energy range, no fewer than 30 energy points were tested, with measurements taken at 0.1 keV steps near the absorption edges of LaBr$_3$(Ce,Sr) crystal (13.47 keV and 38.93 keV) and NaI(Tl) crystal (33.17 keV).
% 本研究采用的是已封装好的1英寸圆柱形LaBr3(Ce,Sr)和NaI(Tl)晶体样品。其入射窗由0.2mm厚的铍片制成，四周被1.5mm厚的铝合金外壳包裹，光输出窗口配备了3mm厚的石英玻璃。入射面和柱面涂有0.4mm厚的特氟龙(Teflon)反射材料，用于反射闪烁光子，从而提高光子的收集效率(图2（a）)。通过硅油将待测试的晶体与光电倍增管（PMT）耦合，PMT是滨松公司的CR160型号。本研究中测试LaBr3(Ce,Sr)晶体时PMT工作电压为-800V，测试NaI(Tl)晶体时为-1000V。图2（b）是被测晶体探测器的数据采集系统框图。X射线束流从铍窗入射，与晶体发生相互作用并产生闪烁光子，闪烁光子在PMT光阴极处被收集并转换为光电子，经打拿极倍增后产生一个电荷脉冲信号。我们使用Flash ADC DT5751数字化仪并在实验室的电脑上用LABVIEW记录信号，然后通过网线连接到监控室的电脑上以实现控制数据采集。实验环境温度控制在20+-1℃。对于10-100keV能量范围内，测试能点不少于30个，并且在LaBr3(Ce,Sr)晶体的吸收边（13.47keV和38.93keV）和NaI(Tl)晶体的吸收边（33.17keV）附近都以0.1keV为步长测试。

Both LaBr$_3$(Ce,Sr) and NaI(Tl) crystals exhibit non-uniformity, which is a significant factor affecting energy resolution. The position of X-ray incidence varies, leading to differences in the locations where scintillation photons are generated within the crystals. These photons undergo refraction, reflection, or absorption, resulting in a varying collection efficiency at the PMT photocathode, as only a portion of the photons reach it. To measure the position response of the crystals, we tested 25 locations on both crystals using 25 keV X-rays from the HXCF single-crystal testing system. These positions are distributed as shown in Fig.~\ref{fig:2} (a), with the beryllium window as a reference plane ($XY$-plane), and the 25 points evenly arranged along both the $X$ and $Y$ directions. Similar to the procedure for testing energy response, we first adjusted the monochromator angle and collected 100 seconds of energy spectrum data with an HPGe detector to determine the X-ray energy and flux. We then set the appropriate coordinates and moved each position sequentially to the X-ray exit, ensuring that the X-rays precisely hit the designated location. During the testing of each position, we ensured that the count exceeded 30,000, and background data were collected both before and after the experiment.
% LaBr3(Ce,Sr)和NaI(Tl)晶体都存在不均匀性，这也是导致能量分辨率的一个不可忽视的因素。X射线入射位置不同，晶体内部闪烁光子产生位置也不同。这些光子将遭遇折射、反射或吸收等过程，最终只有部分光子到达光阴极处，造成PMT光阴极处收集效率的差异。为了测量晶体的位置响应，我们用来自HXCF单晶测试装置的25keV X射线测试了这两种晶体上的25个位置点。这些位置点分布如图2（a）所示，以铍窗为一个平面，25个点沿着X和Y两个方向均匀分布。与测试能量响应的操作相似，我们先调节晶体单色仪的角度，用HPGe探测器采集100s能谱数据以确定X射线的能量和通量。然后我们设置了正确的坐标，控制平台依次将每个位置点移动到X射线出口，保证X射线刚好入射到相应的位置。测试每个位置点时，需要保证计数大于30000个，且在实验前、后各采集一次本底数据。

\subsection{PMT Single-photoelectron Calibration}

Photoelectron statistical fluctuations are one of the key factors affecting energy resolution. To investigate the contribution of this component, it is necessary to calibrate the PMT and calculate the photoelectron yield. We used an LED driven by pulsed voltage to illuminate the CR160 PMT and measure the photoelectron spectrum. From this spectrum, the peak position and resolution of the PMT for the single photoelectron were obtained. The single-photoelectron resolution is a critical indicator of PMT performance and a significant contributor to the energy resolution of the crystal detectors, which is the focus of this study. This LED-based method, involving the actual photoelectric conversion process, allows control of single and multiple photoelectron production by adjusting the LED intensity. It is widely regarded as an effective technique for calibrating PMTs \cite{wei2018consistency}.
% 光电子统计涨落是影响能量分辨率的重要因素之一，为了研究该成分的贡献量，需要刻度PMT，并计算光电子产额。我们采用基于脉冲电压驱动的LED来照射CR160 PMT，以测量光电子谱。由单光电子谱可以获得PMT对单光电子的峰位和分辨率。单光电子分辨率是评价PMT性能的重要指标，也是本研究关注的晶体能量分辨率的一个重要来源。这种方法涉及真实的光电转换过程，通过调节LED光源的强弱可以控制单光电子和多光电子的产额，被认为是刻度PMT的有效方法而广泛应用。

Figure~\ref{fig:2} (c) shows the PMT single-photoelectron calibration setup based on LED illumination driven by a pulsed voltage. The red and blue markers indicate the PMT's photoelectron signal and the external trigger signal of the data acquisition system, respectively. To ensure the accuracy and reliability of the experiment, the PMT had to be placed in a light-tight dark box, and a -1600 V operating voltage was supplied by a power supply (model N1470
from CAEN). The pulse generator was used to drive the LED and provided a positive polarity synchronization signal, which was converted into a negative polarity signal via the fan-in/fan-out module before being sent to the gate generator to adjust the gate width. This gate signal utilized as an
external trigger for the digitizer (model DT5751 from CAEN), effectively eliminating noise interference. First, we adjusted the LED voltage and monitored the digital oscilloscope to maintain the probability of the PMT photoelectron signal appearing within the gate signal range at an appropriate level (approximately 1/10). Using this voltage as a reference, we further fine-tuned the LED voltage amplitude and pulse width to stabilize the PMT signal position recorded by the DT5751, ensuring that it remained unchanged regardless of changes in the LED pulse parameters. At this point, the PMT output signal was considered to be at the single-photon level. In this state, we collected data from more than 30,000 events and subsequently plotted the photoelectron spectrum of the PMT operating at -1600 V for further analysis and research.
% 图2（c）是基于脉冲电压驱动LED照射的PMT单光电子刻度装置，红色和蓝色标记分别表示PMT的光电子信号和数据采集设备的外部触发信号。为了保证实验的准确性和可靠性，我们将PMT放在暗箱中做严格的避光处理，并由插件CAEN N1470提供-1600V的最高工作电压。脉冲信号发生器RIGOL DG1062Z驱动LED发光，并提供一路正极性同步信号，接入扇入扇出模块CAEN N625后被转换为负极性，再接入GG8000调整其门宽。该门信号作为数据采集设备DT5751的外部触发，以有效消除噪声的干扰。首先，调节LED驱动电压，观察数字示波器Tektronix DPO3054，将PMT光电子信号出现在门信号之内的概率保持在合适的范围（大约1/10）。以此电压值为中心，再调节LED脉冲电压幅度和脉冲宽度，使DT5751所记录的PMT信号位置基本固定，不随LED驱动脉冲参数的改变而发生明显移动，此时认为PMT输出信号处于单光子水平。在该状态下，采集大于30000个事例的实验数据，进而绘制出PMT工作在-1600V时的光电子谱，用于后续分析研究。

When the PMT operates at low voltage, it fails to generate sufficient gain to detect single-photon signals. To measure the absolute light yield of LaBr$_3$(Ce,Sr) and NaI(Tl) crystals, it is necessary to indirectly assess the single-photoelectron response of the PMT at -800 V and -1000 V. We used the PMT to test the energy spectrum of the LaBr$_3$(Ce,Sr) crystal with a $^{241}$Am radioactive source. During the test, the positions of the radioactive source and the LaBr$_3$(Ce,Sr) crystal were held constant while the PMT operating voltage was sequentially set to -800 V, -1000 V, and -1600 V for data collection. A fitting analysis of the energy spectra at these three voltages yielded the full-energy peak positions of the 59.5 keV line from the $^{241}$Am source. Under unchanged experimental conditions, we assume that the number of photons received at the PMT photocathode remains relatively constant. The single-photoelectron response at -800 V and -1000 V can be calculated using Equation~\ref{eq:3}. This method addresses the issue of insufficient PMT gain, allowing for calibration at low operating voltages.
% 当PMT工作在低电压时无法产生足够的增益以探测到单个光子的信号。为了测得LaBr3（Ce,Sr）和NaI(Tl)晶体的绝对光产额，需要间接测量PMT工作在-800V和-1000V时的单光电子响应。我们使用该PMT测试了LaBr3（Ce,Sr）晶体对241Am放射源的能谱。测试时，保持放射源和LaBr3（Ce,Sr）晶体的位置不变，PMT工作电压依次设置为-800V、-1000V和-1600V，并进行数据采集。拟合分析这三个电压下的能谱，得到241Am放射源的59.5keV全能峰峰位。在保证实验条件不变的情况下，认为PMT光阴极处接收的光子数目是基本不变的。根据公式3可以计算出PMT工作在-800V和-1000V时的单光电子响应。该方法解决了PMT增益不足的问题，实现了PMT低工作电压时的刻度。

\begin{equation}\label{eq:3}
N = \frac{ADC_{-1600 V}}{ADC_{spe, -1600 V}} = \frac{ADC_{-1000 V}}{ADC_{spe, -1000 V}} = \frac{ADC_{-800 V}}{ADC_{spe, -800 V}} .
\end{equation}

In Equation~\ref{eq:3}, $N$ denotes the number of photoelectrons at the PMT photocathode. 
$ADC_{-1600 V}$, $ADC_{-1000 V}$, and $ADC_{-800 V}$ correspond to the peak positions of the LaBr$_3$(Ce,Sr) crystal at 59.5 keV when the PMT operates at -1600 V, -1000 V, and -800 V, respectively. $ADC_{spe, -1600 V}$ represents the single-photoelectron response of the PMT at -1600 V, obtained using the LED-based PMT calibration method. $ADC_{spe, -1000 V}$ and $ADC_{spe, -800 V}$ refer to the single-photoelectron responses at -1000 V and -800 V, respectively, which are the primary focus of this study.
% 其中，N表示PMT光阴极处光电子数，ADC_{-1600 V}、ADC_{-1000 V}和ADC_{-800 V}分别表示PMT工作在-1600 V、-1000 V和-800 V时LaBr$_3$(Ce,Sr)晶体的59.5keV峰位。ADC_{spe, -1600 V}表示PMT工作在-1600 V时的单光电子响应，由基于LED的PMT刻度法获得。ADC_{spe, -1000 V}和ADC_{spe, -800 V}则分别表示PMT工作在-1000 V和-800 V时的单光电子响应，是本研究关注的重点。

\section{Monte Carlo simulation}

Geant4 is a Monte Carlo (MC) simulation software package developed by the European Organization for Nuclear Research (CERN) to model the transport of particles through matter \cite{lu2022monte, huyan2018geant4}. In this study, a scintillator detector model identical to that used in the HXCF experiment described in Section~\ref{chap:Hard X-ray Ground Calibration Facility} was established using Geant4 version 10.2. 
% Geant4是由欧洲核子中心(CERN)开发的蒙特卡洛（MC）应用软件包，用于模拟粒子在物质中的输运过程。本研究采用Geant4 10.2版本建立了与2.1章节HXCF实验中完全相同的闪烁体探测器模型.

\subsection{Energy Response Distribution}\label{chap:Energy Response Distribution}

We selected the $G4EmLivermorePhysics$ model to handle the interactions of low-energy gamma photons with matter. The energy deposition process of a 100 keV gamma-ray beam was simulated in the two types of crystals, with 30,000 incident events for each simulation. Only a portion of the events resulted in full-energy deposition within the crystals. Gamma rays interact with the crystal to generate secondary electrons through two mechanisms: (1) the direct photoelectric effect, which produces multiple Auger electrons and characteristic X-rays. These characteristic X-rays are reabsorbed by the crystal, generating secondary photoelectrons; and (2) Compton scattering, which produces one or more Compton electrons, followed by the secondary gamma photons undergoing a photoelectric effect cascade \cite{FPY}.
% 我们选择了G4EmLivermorePhysics模型来处理低能伽马光子与物质的相互作用。模拟100keV伽马射线束分别在两种晶体中的能量沉积过程，每次束流有30000个入射事例。只有部分事件全部能量都沉积在晶体中。伽马射线与晶体发生相互作用产生次级电子的方式有两种：（1）直接发生光电效应级联过程产生多个俄歇电子和特征X射线，特征X射线被晶体再吸收后产生次级光电子；（2）康普顿散射后产生一个或几个康普顿电子，而次级光子将发生光电效应级联过程。

We selected only full-energy deposition events, as these are the focus of this study, and obtained the energies of secondary electrons generated during the material interactions. Notably, we specifically selected Compton electrons, photoelectrons, and Auger electrons (collectively referred to as secondary electrons), excluding any electrons originating from ionization or other processes. By convolving the energies of these secondary electrons with the non-proportional electron response curves of the LaBr$_3$(Ce,Sr) and NaI(Tl) crystals \cite{FPY}, we reconstructed the energy response distribution for the full-energy deposition events. The spread of this distribution, represented by the standard deviation ($\sigma$), can be used to assess the contribution of non-proportionality to the energy resolution.
% 我们挑选出全能量沉积事件，只有这些全能量沉积事件才是本研究的关注点，并获取物质相互作用过程中次级电子的能量。特别注意的是，只挑选康普顿电子、光电子和俄歇电子（三者统称次级电子），不能挑选来自电离或其他方式的电子。用这些次级电子能量卷积LaBr$_3$(Ce,Sr)和NaI（Tl）晶体对电子的光产额非线性曲线，可以重建出全能量沉积事件的能量响应分布。这一分布的离散程度(用标准差σ表示)可用于衡量非线性对能量分辨率的贡献。

\begin{equation}\label{eq:4}
L_{\gamma,i} = \sum_{j=1}^{M_i}R_e(E_{e,j})E_{e,j} .
\end{equation}

\begin{equation}\label{eq:5}
R_\gamma = \frac{\overline{L_\gamma}}{E_\gamma} = \frac{1}{E_\gamma} \frac{1}{N} \sum_{i=1}^NL_{\gamma,i} .
\end{equation}

\begin{equation}\label{eq:6}
\sigma^2 = \frac{1}{N-1}\sum_{i=1}^N(L_{\gamma,i}-\overline{L_\gamma})^2 .
\end{equation}

\begin{equation}\label{eq:7}
\delta_{non} = \frac{\sigma}{\overline{L_\gamma}} .
\end{equation}

Equation~\ref{eq:4} represents the convolution operation involved in this theoretical method, where $L_{\gamma,i}$ denotes the light yield of the $i$-th gamma particle, $M_i$ is the number of secondary electrons generated by the $i$-th gamma particle, and $E_{e,j}$ represents the energy of the $j$-th secondary electron. Here, $R_e(E_{e,j})$ denotes the relative light yield of the $j$-th secondary electron, obtained from the non-proportional electron response curves of the crystals. The non-proportional electron response curves used in this study is based on our previously published results \cite{FPY}. The aforementioned convolution operation is applied to $N$ full-energy deposition gamma events, yielding an energy response distribution $L_\gamma$. Equation~\ref{eq:5} defines the calculated response $R_\gamma$ as the ratio of the mean value of this distribution, $\overline{L_\gamma}$, to the incident gamma particle energy $E_\gamma$. The standard deviation $\sigma$, representing the spread of the energy response distribution, is given by Equation~\ref{eq:6}. The ratio of $\sigma$ to the mean $\overline{L_\gamma}$, denoted by $\delta_{non}$, quantifies the contribution of the crystals' non-proportionality to the total energy resolution (Equation~\ref{eq:7}) \cite{valentine1998light}.
% 公式1表示该理论方法涉及的卷积运算，其中L_{\gamma,i}表示第i个伽马粒子的光产额，M_i是第i个伽马粒子产生的次级电子数，E_(e，j)表示第j个次级电子的能量。在这里，R_e(E_{e,j})表示第j个电子的相对光产额，由晶体对电子的光产额非线性曲线获得。本研究采用的电子光产额非线性曲线是我们已经公布的结果[FPY_article1]。对N个全能量沉积的伽马粒子事件，都进行上述卷积运算，可以得到一个能量响应分布L_\gamma。公式2定义计算响应R_\gamma为该分布的均值——L_\gamma与入射的伽马粒子能量E_\gamma的比值。σ表示能量响应分布的标准差（公式3），它与均值——L_\gamma的比值δ_(non)，便是晶体非线性对总能量分辨率的贡献量（公式4）。

\subsection{Photon Collection Efficiency}

To investigate the contribution of photon collection non-uniformity to the total energy resolution, we simulated the transport and collection processes of scintillation photons in the LaBr$_3$(Ce,Sr) and NaI(Tl) crystals when photons are emitted from different positions. These scintillation photons undergo reflection, refraction, or absorption by materials such as Teflon reflective films, silicone grease, and light guides. This process affects the collection efficiency at the PMT photocathode, which in turn impacts the energy resolution. The performance parameters of the crystals are listed in Table~\ref{tab:1}. In the Geant4 optical simulation, we set the optical properties of the crystal and used the $G4OpticalSurface$ class to describe the boundary optical properties, setting the boundary type to $dielectric\_dielectric$, the surface finish to $ground$, and the boundary calculation model to $unified$. The material properties involved in the detector model are shown in Table~\ref{tab:2}. In the $G4OpticalPhysics$ class, optical processes such as scintillation photon generation, Cherenkov radiation, Burke absorption, Rayleigh scattering, and various boundary processes (reflection, refraction, and absorption) were set.
% 为了研究光子收集不均匀性对总能量分辨率的贡献量，我们还模拟了NaI（Tl）晶体在不同位置发光时闪烁光子的输运及收集过程。这些闪烁光子被特氟龙反射膜、硅油、光导等材料反射、折射或吸收，这一过程影响着PMT光阴极处的收集效率，也会造成能量分辨率上的差异。LaBr3(Ce,Sr)和NaI（Tl）晶体的性能参数如表1所示。Geant4光学模拟时，我们设置好晶体的光学属性，并使用G4OpticalSurface类来描述边界光学属性，选择合适的边界类型、边界计算模型和边界表面处理。探测器模型涉及到的材料属性如表2所示。在G4OpticalPhysics类中设置光学过程，包括闪烁光子产生、切利科夫辐射、伯克吸收、瑞利散射和一些边界过程（反射、折射和吸收）。

After setting up the series of parameters, properties, and optical processes, we simulated scintillation photon generation at five depths ($Z$-axis, 1 mm, 2 mm, 3 mm, 4 mm, and 5 mm) within the crystals. At each depth, a point light source emitted 20,000 scintillation photons isotropically in a 4$\pi$ solid angle from 13 different radial positions along the crystals. We tracked the transport of scintillation photons within the detector model and recorded the number of photons reaching the PMT window. This study does not consider the PMT's quantum efficiency or subsequent multiplication processes. The photon collection efficiency is defined as the ratio of the number of scintillation photons reaching the PMT window to the number of photons generated at the scintillation center within the crystals. The variation in collection efficiency was used to evaluate the contribution of non-uniformity to the total energy resolution.
% 在设置好这一系列参数、属性和光学物理过程后，我们模拟在晶体内1 mm、2 mm、3 mm、4 mm和5 mm五种轴向深度下产生闪烁光子的情况。每个深度下，在沿着晶体径向的13个位置，点光源以4π立体角发射20000个闪烁光子。我们跟踪了闪烁光子在探测器模型中的输运过程，并统计了能到达PMT端窗的闪烁光子数。本研究不讨论PMT的量子效率及后续的倍增过程，光子收集效率定义为PMT端窗处的闪烁光子数与晶体内部发光中心产生闪烁光子数之比。根据收集效率的差异来衡量不均匀性在总能量分辨率中的贡献量。

\begin{table}
\caption{Performance parameters of the LaBr$_3$(Ce,Sr) and NaI(Tl) crystals set in the Geant4 simulation.}
\label{tab:1}
\setlength{\tabcolsep}{3pt}
\begin{tabular}{|p{105pt}|p{66pt}|p{66pt}|}
\hline
Item & LaBr$_3$(Ce,Sr) & NaI(Tl)  \\
\hline
Density & 5.10 g/cm$^3$ & 3.67 g/cm$^3$\\
Refractive index & 1.90 & 1.85 \\
Light yield & 63000 photons/MeV & 38000 photons/MeV \\
Intrinsic resolution & 1 & 1 \\
Maximum emission wavelength & 380 nm & 415 nm \\
Decay time & 20 ns & 230 ns \\
\hline
\end{tabular}
\end{table}

\begin{table}
\caption{Materials properties involved in crystal detector models.}
\label{tab:2}
\setlength{\tabcolsep}{3pt}
\begin{tabular}{|p{70pt}|p{60pt}|p{60pt}|p{40pt}|}
\hline
Item & Material & Refractive index & Reflectivity \\
\hline
Coupling material & Silicone grease & 1.40 & / \\
Optical window & Quartz glass & 1.47 & / \\
PMT window & Borosilicate glass & 1.54 & / \\
Reflective material & Teflon film & 1.35 & 95\% \\
\hline
\end{tabular}
\end{table}

\section{Results}

This section presents the energy and spatial responses of the crystal detectors, the results of PMT single-photoelectron calibration, and the outcomes of Geant4 simulations. We discuss seven factors affecting energy resolution: (1) fluctuations in the energy transfer process, (2) non-proportional luminescence of the crystal, (3) non-uniform photon collection at the PMT photocathode, (4) statistical fluctuations in the conversion of photons to photoelectrons, (5) the single-photoelectron resolution of the PMT, (6) electronic noise, and (7) temperature drift. In this study, we utilized the high-performance PMT, allowing us to neglect the contribution of dark noise. The ambient temperature was maintained at 20±1°C, thus minimizing the impact of temperature drift. The contributions of the remaining five factors to the total energy resolution are detailed in this section.
% 本章节展示了晶体探测器能量响应和位置响应、PMT单光电子刻度以及Genat4模拟的结果。我们讨论了六种影响能量分辨率的因素，分别是：（1）能量传递过程的涨落、（2）晶体自身非线性的发光、（3）PMT光阴极处对光子的收集具有不均匀性、（4）光子转换为光电子过程中统计涨落的贡献、（5）光电倍增管的单光电子分辨率、（6）电子学噪声、（7）温度漂移。在本工作中，我们使用了性能优良的PMT，因此暗噪声的贡献可以忽略不计。实验中环境温度保持在20+-1℃，因此温度漂移的贡献也可以忽略不计。其余五项因素在总能量分辨率中的贡献量均在本章节中一一展示出来。

\subsection{Energy Response} \label{chap:Energy Response}

\begin{figure*}[!htb]
\centering
\includegraphics
  [width=\hsize]
  {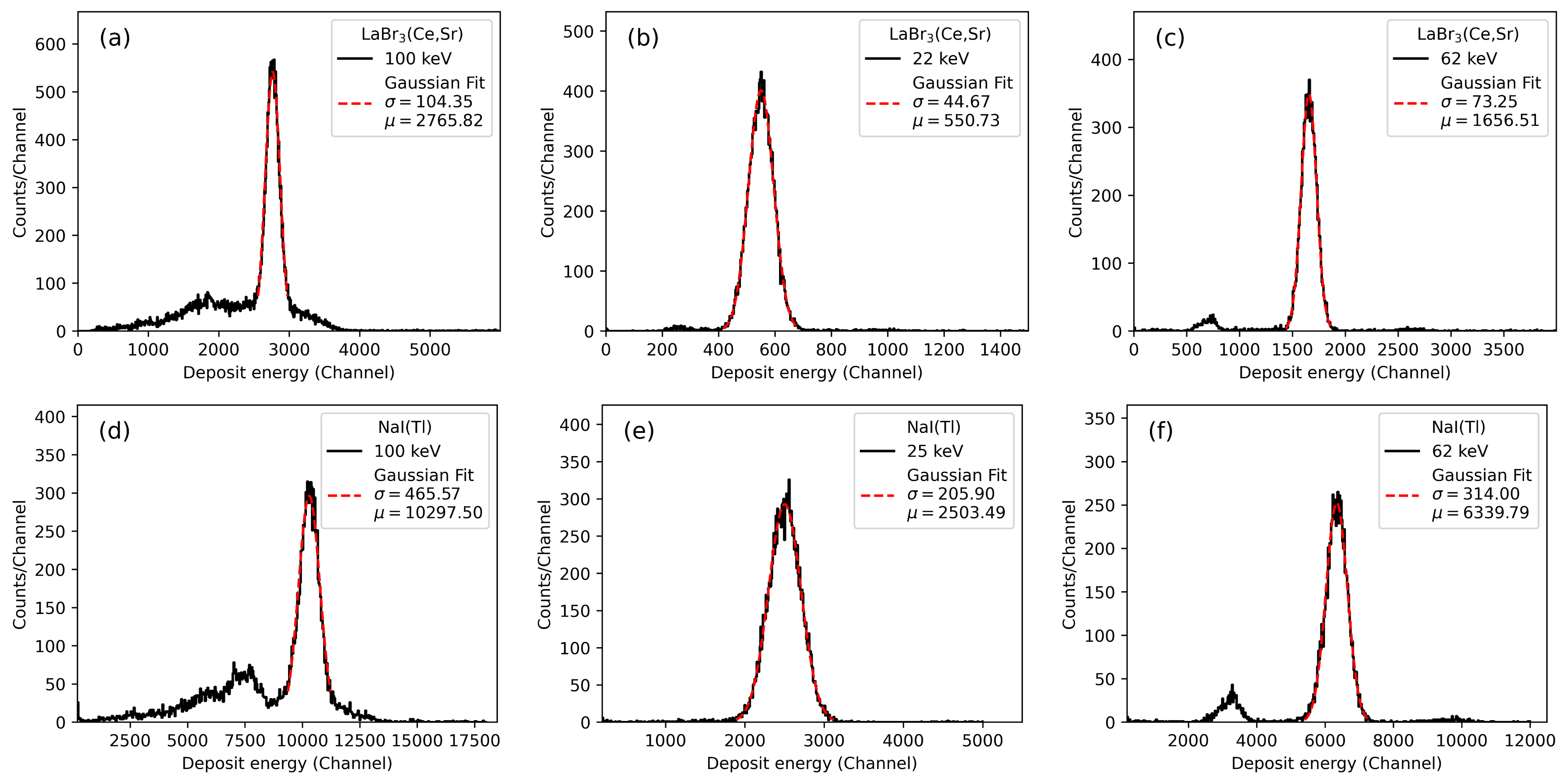}
\caption{Top: Net energy spectra of the LaBr$_3$(Ce,Sr) detector were obtained using X-rays with energies of 100 keV (a), 22 keV (b), and 62 keV (c). Bottom: Net energy spectra of the NaI(Tl) detector were obtained using X-rays with energies of 100 keV (d), 25 keV (e), and 62 keV (f). Background subtraction and Gaussian fitting were performed on each energy spectrum, with the fitting results also presented in the figures.}
\label{fig:13}
\end{figure*}

We used two HXCFs to test the energy response of the LaBr$_3$(Ce,Sr) and NaI(Tl) detectors to 10-100 keV X-rays. Since the background varies with changes in the testing environment, subtracting the background from the X-ray spectrum is essential. Due to the varying test durations, we normalized both the X-ray and background spectra, subtracted the background from the X-ray spectrum, and then denormalized the background-subtracted spectrum to obtain the filtered X-ray spectrum. The filtered X-ray spectrum obtained through this process is referred to as the net energy spectrum. Figure~\ref{fig:13} illustrates the partial net energy spectra and Gaussian fitting results for the LaBr$_3$(Ce,Sr) and NaI(Tl) detectors. These precise fitting results, including the centroid of the energy peak $\mu_i$, standard deviations $\sigma_i$, and full width at half maximum ($FWHM = 2.355 \cdot \sigma_i$), provide a foundation for establishing the energy-channel (E-C) and the energy-resolution relationship for the detectors. As shown in Figs.~\ref{fig:13} (a), (b), and (c), when the X-ray energy exceeds the binding energy of Br's K-shell electrons (13.47 keV) and the remaining energy surpasses the threshold, a faint escape peak appears to the left of the full-energy peak in the spectrum. Additionally, when the energy exceeds the binding energy of La's K-shell electrons (38.93 keV), a pronounced escape peak is observed on the left side of the full-energy peak. Figures~\ref{fig:13} (d), (e), and (f) indicate that when the X-ray energy is below the binding energy of I's K-shell electrons (33.17 keV), only a single full-energy peak is present in the spectrum. However, when the X-ray energy exceeds the binding energy of I's K-shell electrons and the remaining energy surpasses the threshold, an escape peak emerges to the left of the full-energy peak.
% 我们采用计量院（NIM）的两套HXCFs测试了NaI(Tl)探测器对10-100keV X射线的能量响应。本底会随着测试环境的变化而改变，对X射线谱扣除本底是必须的。由于测试时间的不一，我们对X射线能谱和本底谱都归一化后，把本底从X射线能谱中扣除，再对扣除本底后的X射线谱去归一化，得到滤波后的X射线谱。这样处理得到的滤波后的X射线谱被称为净能谱。图3和图4分别展示了LaBr3(Ce,Sr)和NaI(Tl)探测器的部分净能谱及其高斯拟合的结果。这些精确的拟合结果，如：能量峰的中心值μi、标准偏差 σi、全宽半高 (FWHM = 2.355 · σi)为我们建立探测器的能量-道数（E-C）关系和能量-分辨率关系提供了保障。由图3可见，当X射线能量超过溴的K壳层电子结合能（13.47 keV）且剩余能量超过阈值时，能谱中全能峰的左侧会出现一个微弱的逃逸峰；而能量超过镧的K壳层电子的结合能（38.93 keV）时，全能峰左侧会出现一个明显的逃逸峰。由图4可见，当X射线能量低于碘的K壳层电子的结合能（33.17 keV）时，能谱中只出现单一的全能峰。然而，当X射线能量超过碘的K壳层电子的结合能且剩余能量超过阈值时，在全能峰的左侧会出现一个逃逸峰。

\begin{figure*}[!htb]
\centering
\includegraphics
  [width=0.85\hsize]
  {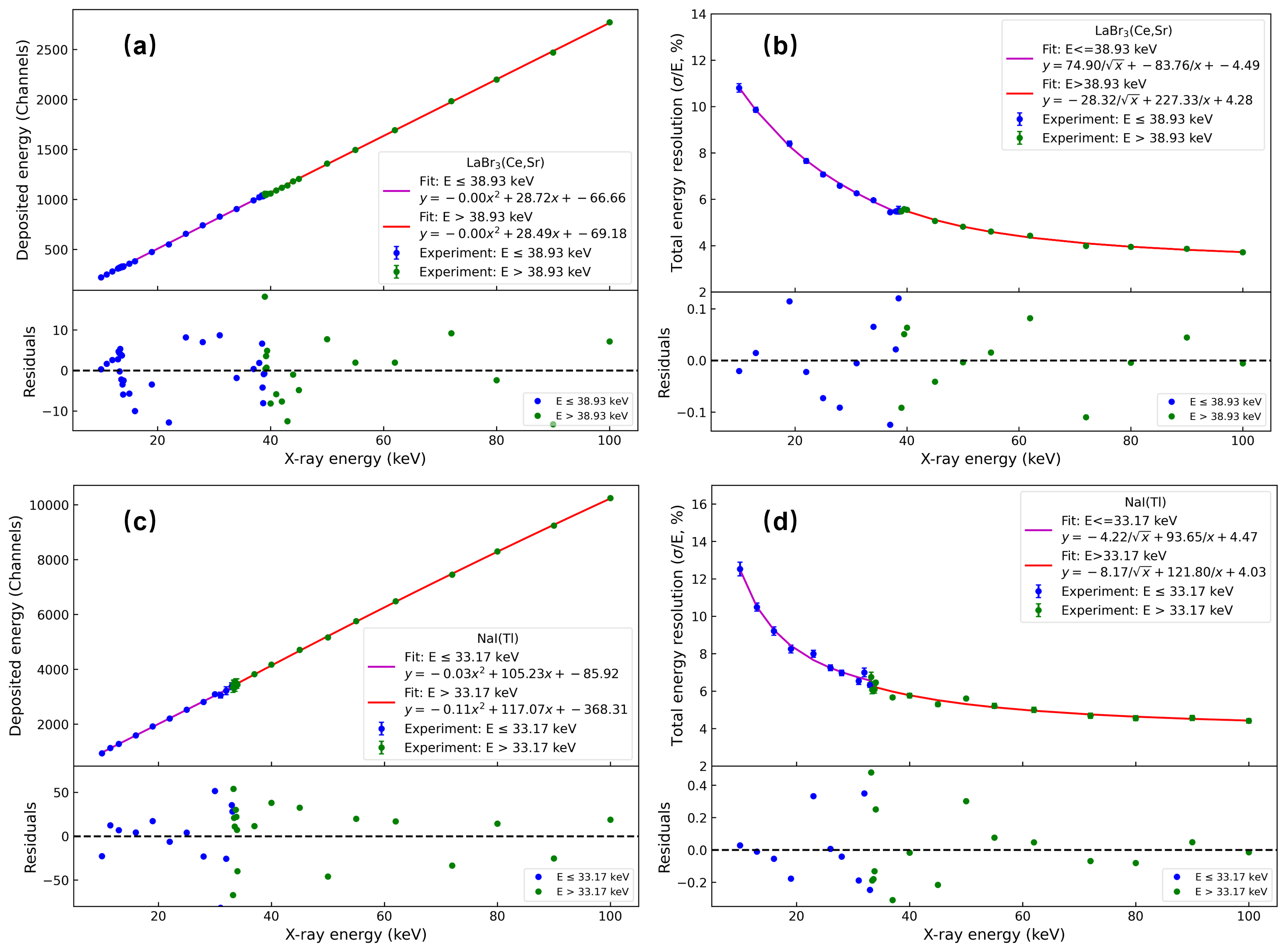}
\caption{Left: Energy-channel relationships of the LaBr$_3$(Ce,Sr) (a) and NaI(Tl) (c) detectors were established by fitting these data points with two quadratic polynomials. Right: Energy-resolution relationships and fitted curves for the LaBr$_3$(Ce,Sr) (b) and NaI(Tl) (d) detectors. The residuals were obtained by subtracting the fitted values from the experimental ones.}
\label{fig:4}
\end{figure*}

It is important to note that the peak positions ($\mu_i$) in the spectrum shown in Fig.~\ref{fig:13} have not yet been corrected for baseline. Therefore, we specifically studied the impact of baseline on the ADC channels acquired by the DT5751 digitizer. The waveforms collected in this study show a baseline fluctuation of approximately 2 mV, which introduces an additional contribution to the ADC channels represented by waveform integration. This contribution is proportional to the waveform width, as expressed in Equation~\ref{eq:8}, where $ADC_{baseline}$ represents the ADC channels introduced by the baseline, and $t_{wavewidth}$ represents the waveform width. This provides an effective method for baseline correction. In fact, the baseline effect can also be observed in the spectrum. A Gaussian peak induced by the baseline is present at the low-energy end of the spectrum shown in Fig.~\ref{fig:13}, though it is not displayed here. The peak position obtained from the Gaussian fit of the baseline peak represents the baseline level, which varies with the energy deposited by X-rays in the NaI(Tl) crystal. This variation occurs because the deposited energy directly affects the signal waveform width. Subtracting the baseline level from the peak position of the X-rays full-energy peak is also a commonly used baseline correction method.
% 需要注意的是，图3能谱的峰位μi还未进行基线修正。为此，我们专门研究了基线对DT5751数字化仪采集到的ADC通道的影响。本工作中采集到的波形显示基线抖动约为2mV，它会给用波形积分表示的ADC通道引入额外贡献。该贡献量与波形宽度成正比，其关系如公式8所示，其中ADC_baseline表示基线引入的ADC道数，t_wavewidth表示波形宽度。这是一种有效的基线修正的方法。实际上，基线效应在能谱上也能观测到。图3所示的能谱在低能端（Channel小于200）存在一个基线引入的高斯峰，只不过这里没有展示出来。对基线峰做高斯拟合得到的峰位就是baseline level，它会随NaI(Tl)晶体中沉积的X射线能量而变化，这是因为沉积能量直接影响信号波形宽度。把baseline level从X射线全能峰的峰位中扣除，这也是一种常用的基线修正方法。

\begin{equation}\label{eq:8}
ADC_{baseline} = 0.2187 \cdot t_{wavewidth} + 0.49 .
\end{equation}

Additionally, the two HXCFs located in different laboratories exhibit differences in systematic errors. After applying baseline correction and systematic error correction to all X-ray spectra, we obtained the energy-channel and energy-resolution relationships shown in Fig.~\ref{fig:4}. Here, the energy resolution is expressed as the ratio of the standard deviation to the peak position. For the LaBr$_3$(Ce,Sr) crystal, the binding energy of La's K-shell electrons (38.93 keV) serves as the boundary point. Using Equations~\ref{eq:9} and~\ref{eq:10}, we can accurately fit the energy-channel and the energy-resolution relationships within the energy ranges of 10-38.93 keV and 38.93-100 keV, respectively. For the NaI(Tl) crystal, the binding energy of I's K-shell electrons (33.17 keV) is used as the boundary point. Similarly, Equations~\ref{eq:9} and~\ref{eq:10} allow for precise fitting of the energy-channel and the energy-resolution relationships in the energy ranges of 10-33.17 keV and 33.17-100 keV, respectively. In Equation~\ref{eq:10}, the constant term $a$ represents electronic noise, the second term $b$ accounts for statistical fluctuations of scintillation photons and photoelectrons, and the third term $c$ reflects the intrinsic contribution of the scintillator, primarily due to the non-proportionality of the light output \cite{feng2024detector, Ground-calibration-GECAM-C, bissaldi2009ground}. The fitting results and residuals are also shown in Fig.~\ref{fig:4}. For 100 keV X-rays, the total energy resolutions ($\sigma/E$) of the LaBr$_3$(Ce,Sr) and NaI(Tl) crystals are 3.71\% ± 0.03\% and 4.41\% ± 0.14\%, respectively.
% 此外，两套位于不同的实验室的HXCFs还存在系统误差的差异。我们对所有X射线能谱进行基线修正和系统误差修正后，得到了图4所示的能量-通道关系和能量-分辨率关系。在这里，能量分辨率用标准偏差sigma与峰位的比值表示。对于LaBr3(Ce,Sr)晶体，以La的K壳层电子的结合能（38.93 keV）作为分界点，用公式9和10可以分别精确地拟合10-38.93keV和38.93-100keV能量范围内的能量-通道关系和能量-分辨率关系。对于NaI(Tl)晶体，以碘的K壳层电子的结合能（33.17 keV）作为分界点，用公式9和10可以分别精确地拟合10-33.17keV和33.17keV-100keV能量范围内的能量-通道关系和能量-分辨率关系。公式10中的常数项a表示电子噪声，第二项b表示闪烁光子和光电子的统计涨落，第三项c反映了闪烁体的本征贡献，主要源于发光的不成比例性。拟合结果和残差也都展示在图4中。对于100keV的X射线，LaBr3(Ce,Sr)晶体和NaI(Tl)晶体的总能量分辨率σ/E分别是3.71%+-0.03%和4.41%+-0.14%。

\begin{equation}\label{eq:9}
Ch(E_\gamma) = a_0 + a_1x + a_2x^2 .
\end{equation}

\begin{equation}\label{eq:10}
Resolution(E_\gamma) = \frac{2.355 \cdot \sigma(E_\gamma)}{Ch(E_\gamma)} = \frac{\sqrt{a^2+b^2E_\gamma+c^2E_\gamma^2}}{E_\gamma} .
\end{equation}

\subsection{Spatial Response}

\begin{figure*}[!htb]
\centering
\includegraphics
  [width=0.8\hsize]
  {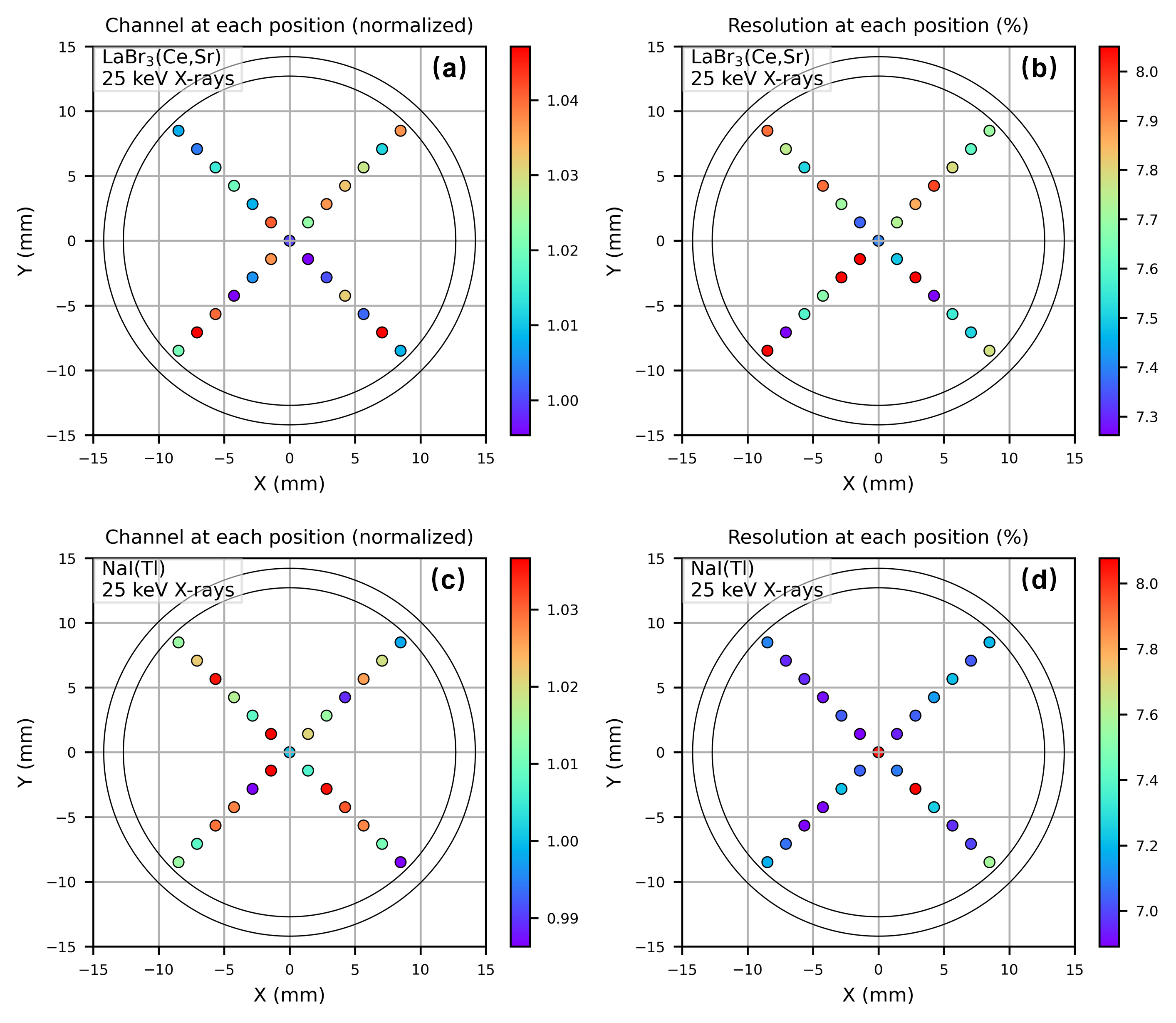}
\caption{Left: Relative full-energy peaks of the LaBr$_3$(Ce,Sr) (a) and NaI(Tl) (c) crystals vary with position in the $X$ and $Y$ directions, with the center of the beryllium window (i.e., the (0, 0) coordinate) defined as "1". Right: Energy resolutions of the LaBr$_3$(Ce,Sr) (b) and NaI(Tl) (d) crystals vary with position in the $X$ and $Y$ directions.}
\label{fig:5}
\end{figure*}

\begin{figure*}[!htb]
\centering
\includegraphics
  [width=0.9\hsize]
  {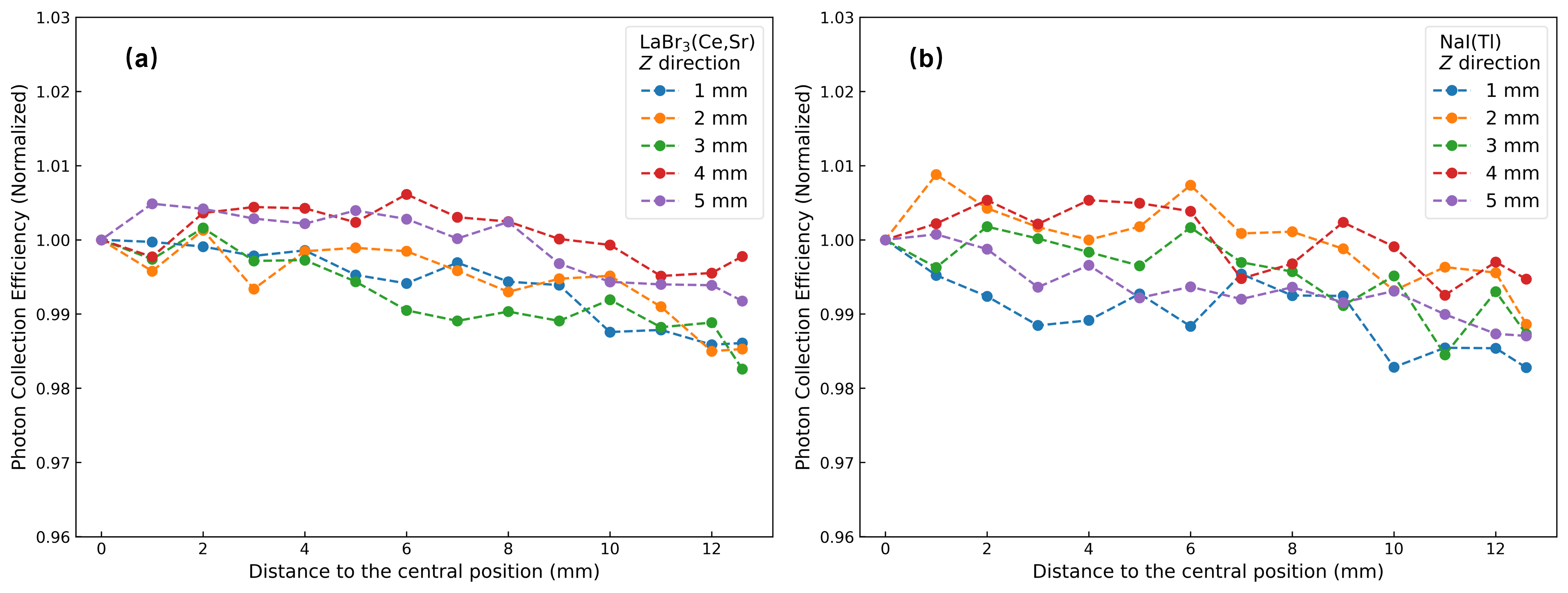}
\caption{Relative photon collection efficiency at the PMT photocathode, obtained from Geant4 simulations, varies depending on the position where the scintillation photons are generated in the LaBr$_3$(Ce,Sr) (a) or NaI(Tl) (b) crystals.}
\label{fig:10}
\end{figure*}

Due to the scintillation non-uniformity of the crystals, the photon collection efficiency at the PMT photocathode varies, resulting in differing responses to X-rays. We used the HXCF to test the position response of the LaBr$_3$(Ce,Sr) and NaI(Tl) detectors to 25 keV X-rays. By setting coordinate parameters for each point, we aligned the X-ray beam with 25 test positions, uniformly distributed in the $X$ and $Y$ directions (with the $Z$ direction as the crystal’s central axis and the $XY$ plane as its cross-section). The position response data were processed using the spectrum fitting and baseline correction methods mentioned in Section~\ref{chap:Energy Response}, yielding the full-energy peak positions and energy resolutions. The normalized relative full-energy peak positions, with the central position set as "1," are shown in Figs.~\ref{fig:5} (a) and (c). The energy resolution is expressed as the ratio of the standard deviation to the peak position, as shown in Figs.~\ref{fig:5} (b) and (d). It was found that for this 1-inch crystals, the peak positions and energy resolutions from the center to the edge do not exhibit a clear monotonic decrease or increase, but rather fluctuate irregularly. We calculated the relative standard deviation ($RSD$) of the peak position distribution using Equation~\ref{eq:11} to evaluate the non-uniformity of the crystal detectors. Calculating the contribution of radial non-uniformity to the energy resolution requires considering the weight of the measurement points' locations \cite{deng2022exploring}. We divided the beryllium window plane ($XY$ plane) in Fig.~\ref{fig:2} (a) into six concentric circles, with radii of 2 mm, 4 mm, 6 mm, 8 mm, 10 mm, and 12 mm, and used the area of each region as the weight for the data points it contains. Based on Equation~\ref{eq:11}, the contribution of radial non-uniformity ($\delta_{un,radial}$) to the energy resolution of the LaBr$_3$(Ce,Sr) and NaI(Tl) detectors are 0.11\% ± 0.01\% and 0.12\% ± 0.01\%, respectively.
% 由于晶体不均匀性，PMT光阴极处的光子收集效率存在差异，这导致对X射线的响应表现出差异。我们使用HXCF测试了LaBr$_3$(Ce,Sr)和NaI(Tl)探测器对25keV X射线的位置响应。通过设置每个点的坐标参数来移动平台将X射线束流依次对准25个待测试位置，这25个位置均匀分布在X和Y两个方向（以晶体中心轴为Z方向，晶体横截面为XY平面）。我们使用4.1章节提到的能谱拟合和基线修正的方法处理了位置响应的数据，得到全能峰的峰位和能量分辨率。以中心位置响应为“1”，归一化后的相对全能峰峰位如图5（a）(c)所示。能量分辨率用标准偏差sigma和峰位的比值表示，如图5（b）(d)所示。可以发现，对于这种1英寸的小型晶体，它从中心到边缘位置的峰位和能量分辨率都不呈现出明显的单调下降或上涨，而是无规律的波动。我们采用公式11计算了峰位分布的相对标准偏差（RSD），以评估NaI(Tl)探测器的不均匀性。计算径向不均匀性对能量分辨率的贡献需要考虑测量点所在位置的权重。我们把图2(a)的铍窗平面（XY平面）划分为6个同心圆，它们的半径分别为2mm，4mm,6mm,8mm,10mm,12mm，并且把每个区域的面积作为它包含数据点的权重。根据公式11，LaBr3(Ce,Sr)和NaI(Tl)探测器的水平不均匀性对能量分辨率的贡献（delta_un,radial ）分别为1.48%+-0.01%和1.57%+-0.01%。

To accurately assess vertical non-uniformity, we performed Geant4 simulations to evaluate the variations in photon collection efficiency at different radial positions of the crystals for depths of 1 mm, 2 mm, 3 mm, 4 mm, and 5 mm along the $Z$ direction, as shown in Fig.~\ref{fig:10}. At a depth of 1 mm along the central axis of the LaBr$_3$(Ce,Sr) and NaI(Tl) crystals, the photon collection efficiencies at the PMT photocathode are 87.73\% and 87.98\%, respectively. All data are normalized to the 1 mm depth at the central axis. From the center of the crystal to the edge, the photon collection efficiency exhibits a fluctuating downward trend, with a relative decrease of no more than 2\%. For positions equidistant from the central axis but at different depths, the relative variation also does not exceed 2\%. After eliminating the effects of radial non-uniformity and simplifying the calculations, we calculated the contributions of vertical non-uniformity ($\delta_{un,vertical}$) to the energy resolution of the LaBr$_3$(Ce,Sr) and NaI(Tl) crystals to be 0.07\% ± 0.01\% and 0.08\% ± 0.01\%, respectively, using Equation~\ref{eq:11} \cite{deng2022exploring}.
% 为了准确衡量垂直不均匀性，我们采用Geant4模拟了1mm、2mm、3mm、4mm和5mm五种深度（Z方向）下沿晶体径向不同位置的光子收集效率变化，结果如图7所示。在LaBr3(Ce,Sr)和NaI（Tl）晶体中心轴1mm深度处产生闪烁光子时，PMT光阴极处的收集效率分别是87.73%和87.98%。所有数据归一化到中心轴1mm深度处。从晶体中心到边缘，光子收集效率呈波动下降趋势，但相对下降不超过2%。对于与中心轴的距离相同，但深度不同时，相对变化也不超过2%。我们消除了水平不均匀性效应并简化计算后，由公式11求得LaBr3(Ce,Sr)和NaI(Tl)晶体垂直不均匀性对能量分辨率的贡献（delta_un,vertical）分别是0.36%+-0.02%和0.42%+-0.02%。

\begin{equation}\label{eq:11}
RSD = \frac{1}{\overline{x}}\sqrt{\frac{1}{n-1}\sum_{i=1}^n (x_i - \overline{x})^2}.
\end{equation}

\subsection{PMT's Single-photoelectron Response}

\begin{figure}[!htb]
\centering
\includegraphics
  [width=\hsize]
  {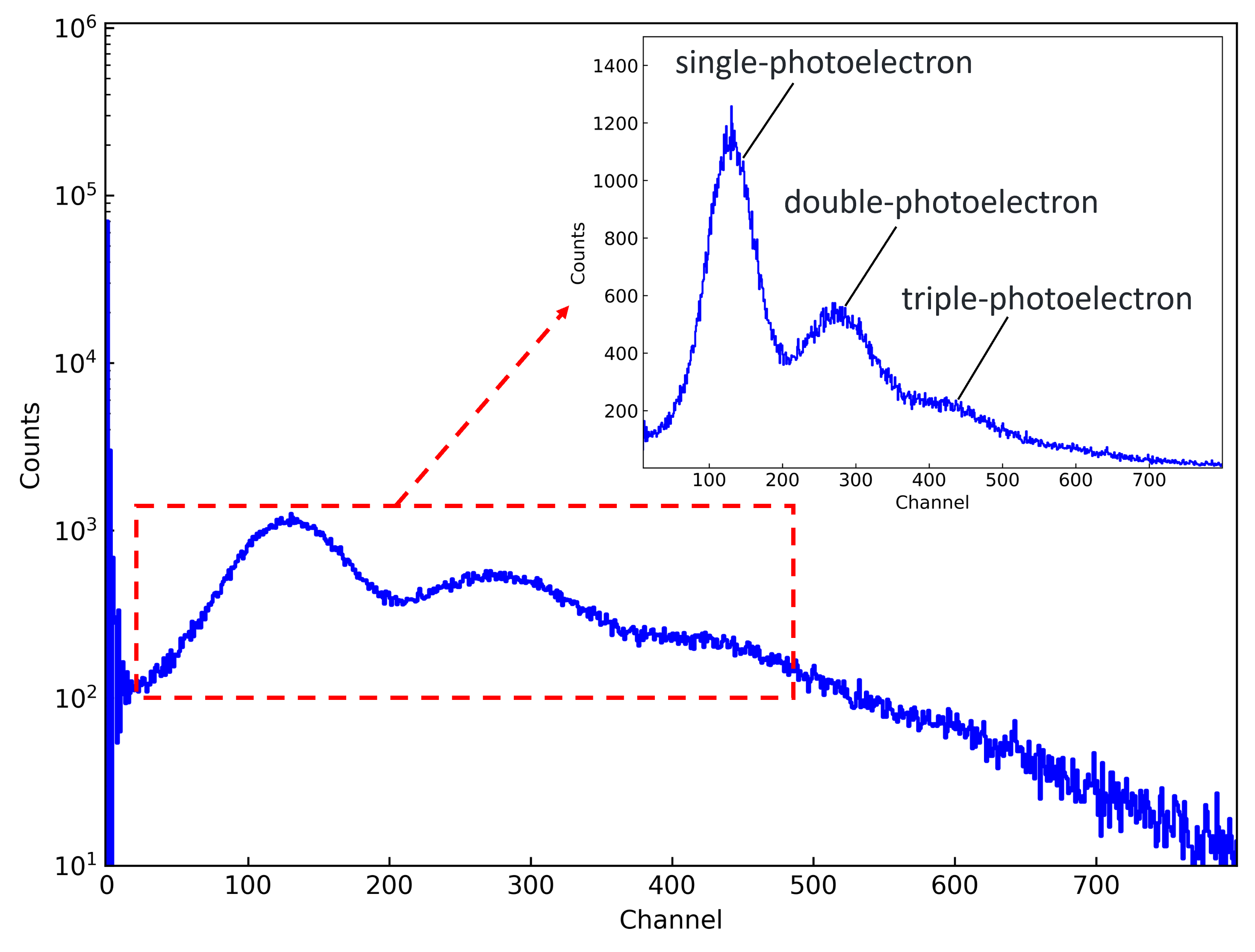}
\caption{Photoelectron spectra measured by the Hamamatsu CR160 PMT, operating at -1600 V.}
\label{fig:6}
\end{figure}

We calibrated the Hamamatsu CR160 PMT for single-photoelectron response using a pulse voltage-driven LED method. By adjusting the amplitude of the pulse voltage, we controlled the intensity of the LED light source to manage the photoelectron yield. Figure~\ref{fig:6} displays the multi-photoelectron spectrum of the PMT operating at -1600 V (note the logarithmic scale on the $y$-axis). A multi-Gaussian fit yielded the single-photoelectron response ($ADC_{spe,-1600 V}$) at -1600 V as 135.10 channels, with a single-photoelectron resolution ($(\sigma/E)_{spe}$) of 27.17\% ± 0.02\%. Theoretically, the operating voltage of the PMT influences its single-photoelectron resolution. However, this study could not measure the PMT single-photoelectron resolution at lower voltages. We believe that the PMT resolution measured at -1600 V is applicable within reasonable error margins to -800 V and -1000 V. To address the gain issue at low operating voltages, we indirectly calibrated the PMT's single-photoelectron response at -800 V and -1000 V by measuring the energy spectrum of the LaBr$_3$(Ce,Sr) crystal using a $^{241}$Am source. After performing a Gaussian fit on the 59.5 keV full-energy peak of the $^{241}$Am source, we calculated the single-photoelectron responses at -800 V and -1000 V to be 2.00 channels and 10.51 channels, respectively, using Equation~\ref{eq:3}. 
% 我们采用基于脉冲电压驱动LED的方法，对滨松CR160 PMT进行了单光电子刻度。通过调节脉冲电压的幅度来控制LED光源的强弱，以实现对光电子产额的控制。图6展示了PMT工作在-1600V下的多光电子谱（注意纵轴为对数坐标）。对其做多高斯拟合，得到PMT工作在-1600V时的单光电子响应ADC_{spe,-1600 V}为135.1 Channels，单光电子分辨率$(\sigma/E)_{spe}$为27.17+-0.02%。从理论上说，PMT的工作电压对其单光电子分辨率有一定的影响。但本研究无法测到更低电压下的PMT单光电子分辨率，我们认为-1600V工作电压下的PMT单光电子分辨率在合理误差内也适用于-800V和-1000V。为了解决低工作电压下增益不足的问题，我们通过测试LaBr3（Ce,Sr）晶体对241Am放射源的能谱来间接刻度-800V和-1000V时的PMT单光电子响应。对241Am放射源的59.5keV全能峰进行高斯拟合后，根据公式3计算出PMT工作在-800V和-1000V时的单光电子响应ADC_{spe,-1000 V}分别是2.00 Channels和10.51 Channels。

\begin{figure*}[!htb]
\centering
\includegraphics
  [width=0.9\hsize]
  {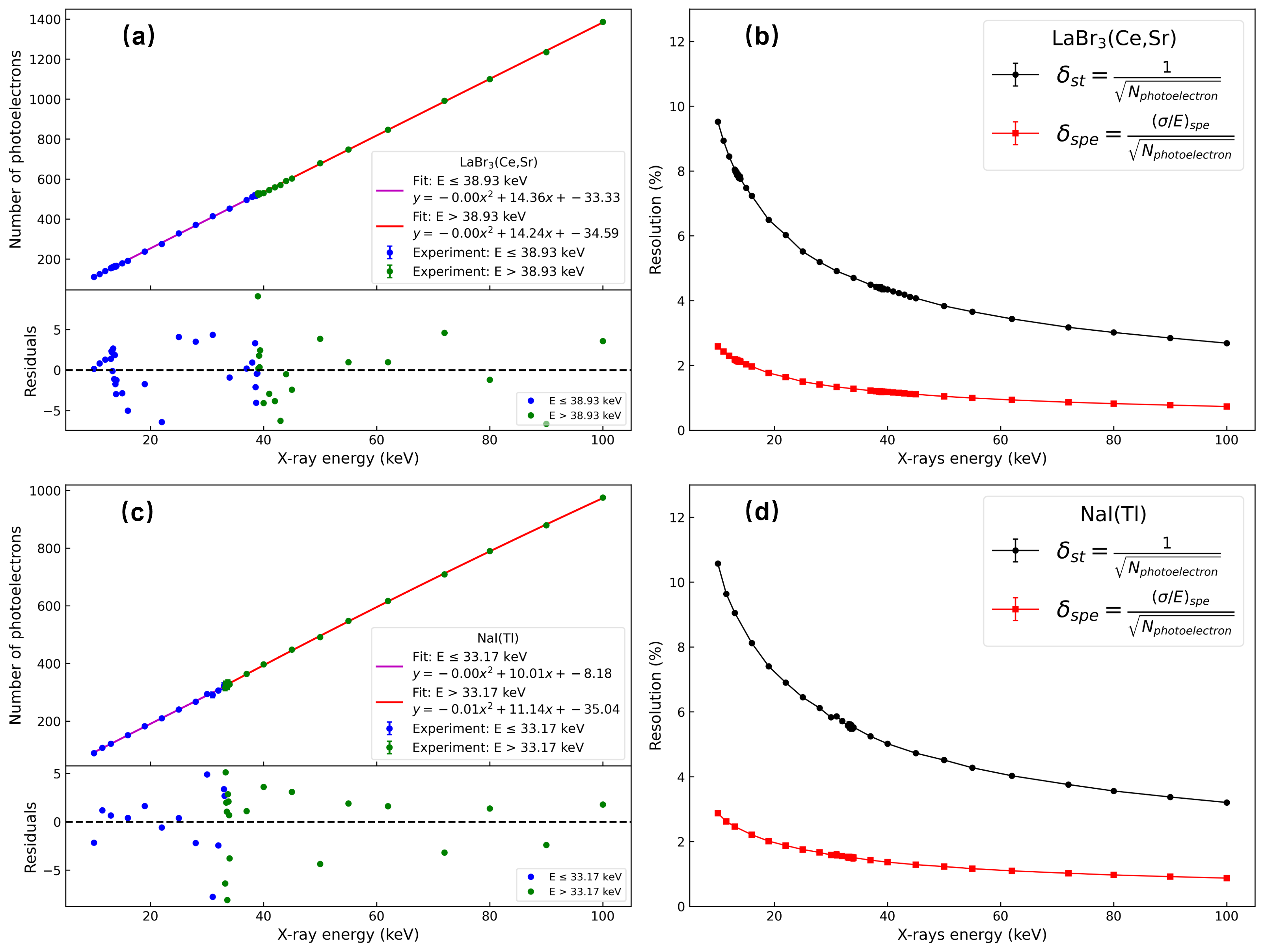}
\caption{Left: Absolute light yield of the LaBr$_3$(Ce,Sr) (a) and NaI(Tl) (c) crystals for X-rays in the 10-100 keV range, expressed in terms of the number of photoelectrons. Right: Contribution of photoelectron statistical fluctuations and PMT single-photoelectron resolution to the total energy resolution of the LaBr$_3$(Ce,Sr) (b) and NaI(Tl) (d) crystals.}
\label{fig:7}
\end{figure*}

In Section~\ref{chap:Energy Response}, we provided the energy-channel ($ADC(E_\gamma)$) relationship for the LaBr$_3$(Ce,Sr) and NaI(Tl) crystals over the 10-100 keV range. Consequently, the absolute light yield of the crystals can be calculated using Equation~\ref{eq:12}, expressed in terms of the number of photoelectrons ($N_{photoelectron}$), with results shown in Figs.~\ref{fig:7} (a) and (c). Using 38.93 keV and 33.17 keV as boundary points, we fitted the data with two quadratic polynomials, and the fitting results are also displayed in Figs.~\ref{fig:7} (a) and (c). The contribution of statistical fluctuations ($\delta_{st}$) in the number of photoelectrons to the total energy resolution is represented by Equation~\ref{eq:13}. We calculated the values of $\delta_{st}$ for the LaBr$_3$(Ce,Sr) and NaI(Tl) crystals in their energy resolutions for 10-100 keV X-rays, as shown in Figs.~\ref{fig:7} (b) and (d). The contribution of the PMT single-photoelectron resolution ($\delta_{spe}$) to the total energy resolution was calculated using Equation~\ref{eq:14}, with results also presented in Figs.~\ref{fig:7} (b) and (d). We observed that statistical fluctuations in photoelectrons significantly affect the total energy resolution. For the LaBr$_3$(Ce,Sr) crystal tested with 100 keV X-rays, 1387 photoelectrons were generated, resulting in $\delta_{st}$ of 2.69\% ± 0.00\% and $\delta_{spe}$ of 0.73\% ± 0.05\%; for the NaI(Tl) crystal, 975 photoelectrons were generated, with $\delta_{st}$ of 3.20\% ± 0.00\% and $\delta_{spe}$ of 0.87\% ± 0.06\%.
% 在章节4.1已经给出了LaBr3(Ce,Sr)和NaI(Tl)晶体对10-100keV X射线的能量-通道(ADC(E_\gamma))关系。因此，根据公式12可以直接计算出晶体的绝对光产额。在这里，绝对光产额可以用光电子数N_{photoelectron}表示，结果如图7（a）和（c）所示。分别以38.93keV和33.17keV为分界点，用两个二次多项式拟合数据，拟合结果也展示在图7（a）和（c）中。光电子数统计涨落对总能量分辨率的贡献δ_st由公式13表示。根据公式13计算出LaBr3(Ce,Sr)和NaI(Tl)晶体对10-100keV X射线能量分辨率中δ_st的值，结果分别展示在图7（b）（d）中。根据公式14计算出PMT单光电子分辨率对总能量分辨率的贡献δ_spe的值，结果也展示在图7（b）和（d）中。我们观察到光电子统计涨落对总能量分辨率的贡献占比较大。用100keV X射线测试LaBr3(Ce,Sr)晶体时，产生了1146个光电子，δ_st为2.95%+-0.01%，并且δ_spe为0.80%+-0.00%；而用100keV X射线测试NaI(Tl)晶体时，产生了975个光电子，δ_st为3.20%+-0.01%，并且δ_spe为0.87%+-0.00%。

\begin{equation}\label{eq:12}
N_{photoelectron}(E_\gamma) = \frac{ADC(E_\gamma)}{ADC_{spe}}.
\end{equation}

\begin{equation}\label{eq:13}
\delta_{st} = \frac{1}{\sqrt{N_{photoelectron}}} .
\end{equation}

\begin{equation}\label{eq:14}
\delta_{spe} = \frac{(\sigma/E)_{spe}}{\sqrt{N_{photoelectron}}} .
\end{equation}

\subsection{Intrinsic Energy Resolution}

\begin{figure*}[!htb]
\centering
\includegraphics
  [width=0.9\hsize]
  {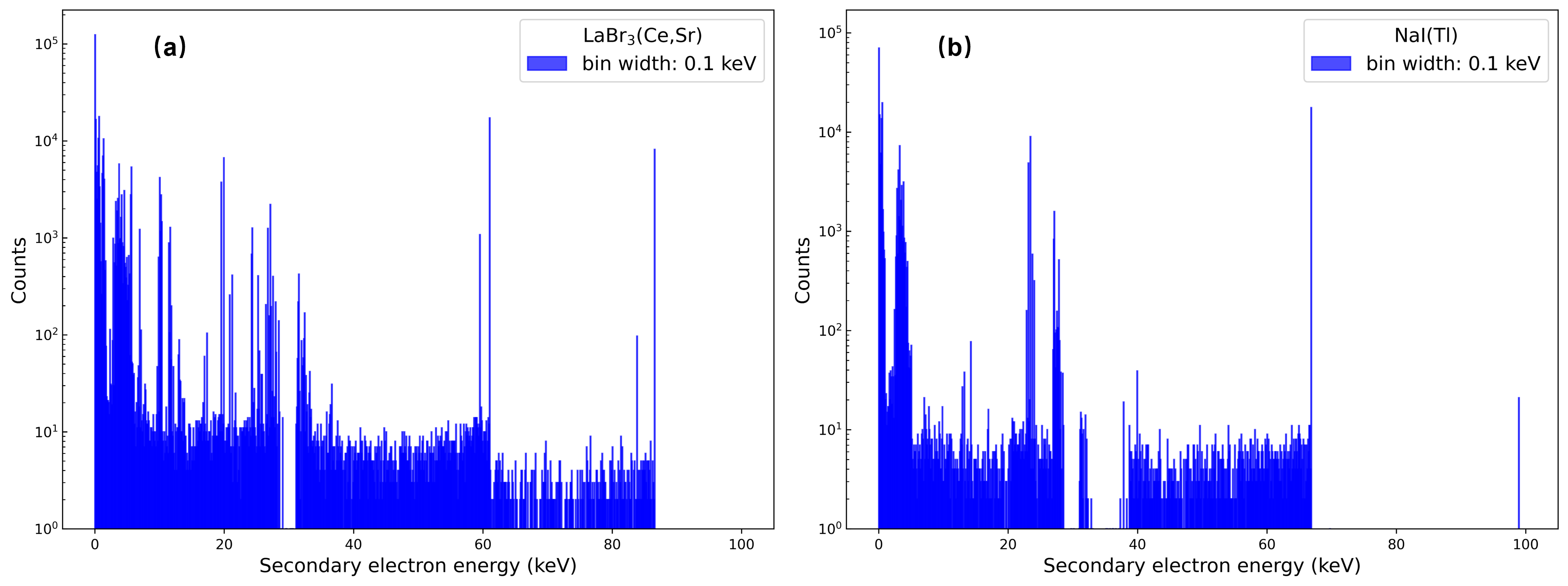}
\caption{Energy distribution of secondary electrons generated by the full-energy deposition of 100 keV gamma photon beams in the LaBr$_3$(Ce,Sr) (a) and NaI(Tl) (b) crystals.}
\label{fig:14}
\end{figure*}

Using the method described in Section~\ref{chap:Energy Response Distribution}, secondary electrons (including Compton electrons, photoelectrons, and Auger electrons) from full-energy deposition events were selected. The energy of these secondary electrons was obtained and convolved with the non-proportional electron response curves of the LaBr$_3$(Ce,Sr) and NaI(Tl) crystals to reconstruct the energy response distributions of the full-energy deposition events. Figures~\ref{fig:14} (a) and (b) illustrate the energy distribution of secondary electrons generated by the full-energy deposition of 100 keV gamma photon beams in LaBr$_3$(Ce,Sr) and NaI(Tl) crystals, respectively. The figures clearly reveal the structures resulting from the absorption edges and atomic energy levels. The continuous spectrum of electron energy is attributed to Compton scattering, while the low-energy electrons in the few keV range are primarily due to a significant number of Auger electrons generated by the photoelectric effect cascade. 
% 根据章节3.1介绍的方法挑选出全能量沉积事件中的次级电子（包括康普顿电子、光电子和俄歇电子）。获取这些次级电子的能量，并与LaBr$_3$(Ce,Sr)和NaI(Tl)晶体对电子的非线性响应曲线做卷积运算，以重建出全能量沉积事件的能量响应分布。图14（a）和（b）分别展示了100keV的伽马光子束的全部能量沉积在NaI(Tl)和LaBr3(Ce,Sr)晶体中时，其产生的次级电子的能量分布图。从图中清晰可见吸收边和原子能级导致的结构。电子能量的连续谱是由于康普顿散射造成的，而几keV的低能电子是由于光电效应级联产生的大量俄歇电子导致的。

\begin{figure}[!htb]
\centering
\includegraphics
  [width=\hsize]
  {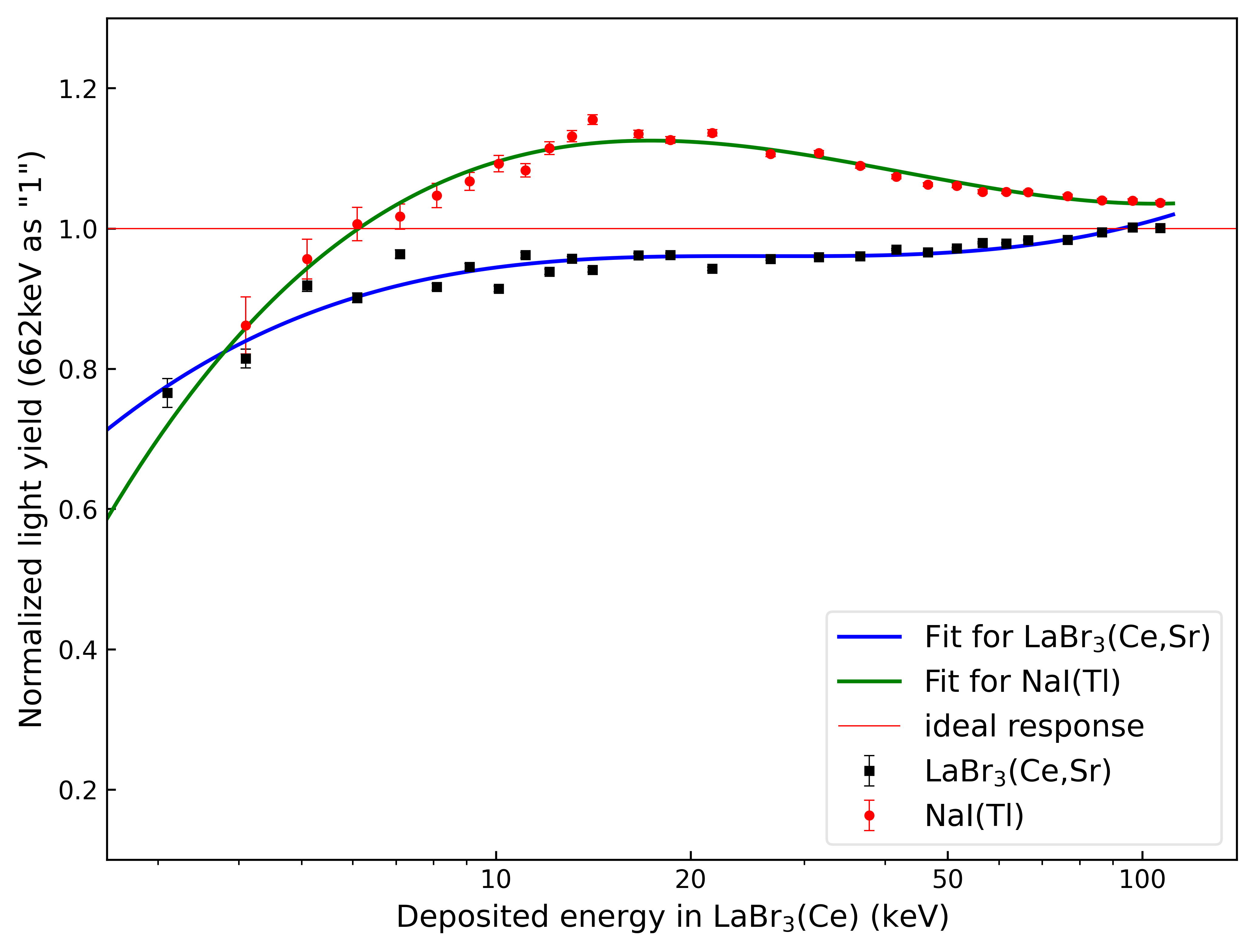}
\caption{Non-proportional electron response curves of LaBr$_3$(Ce,Sr) and NaI(Tl) crystals are referenced from the results of the Wide-Angle Compton Coincidence (WACC) experiments \cite{FPY}.}
\label{fig:15}
\end{figure}

The non-proportional electron response curves used in this study is derived from our published Wide-Angle Compton Coincidence (WACC) experiment results, as shown in Fig.~\ref{fig:15} \cite{FPY}. However, the WACC experiment results provided electron response data for only a limited number of energy points, with the lowest energy point being 3.1 keV. To describe the electron response over a continuous energy range, empirical Equations~\ref{eq:15} and~\ref{eq:16} were used to fit the non-proportional curves \cite{2022Electron}. Here, $NPR(E_e)$ represents the non-proportional response of electrons at different energies $E_e$, and $P_n(n=0,1,2,3)$ are the fitting parameters. Merely fitting the energy range measured in the WACC experiment is insufficient, as Geant4 simulation results indicate that most secondary electrons generated by the photoelectric effect will have energies below a few keV. Therefore, it is necessary to extend the fitting curve to encompass the low-energy range of 0-3.1 keV.
% 本研究采用的非线性曲线是我们已经公布的广角康普顿符合(WACC)实验的结果[FPY_article1]，如图15所示。但WACC实验结果只给出了部分能量点的电子响应，并且最低能量点只能测到3.1keV。为了描述连续能量范围内的电子响应，需要用经验公式15[43]和公式16拟合这条曲线。在这里，NPR(E_e )表示不同能量电子的响应非线性，P_n是拟合参数。仅仅拟合WACC实验中测量到的能量范围是不够的，因为Geant4模拟结果表明大多数由光电效应级联产生的次级电子的能量会低于几keV。因此，有必要将拟合曲线扩展到0–3.1 keV的低能量范围。

\begin{equation}\label{eq:15}
NPR(E_e) = P_0 + P_1 x + P_2 x^2 + P_3 x^3 .
\end{equation}

\begin{equation}\label{eq:16}
x = log(E_e) .
\end{equation}

\begin{figure*}[!htb]
\centering
\includegraphics
  [width=0.9\hsize]
  {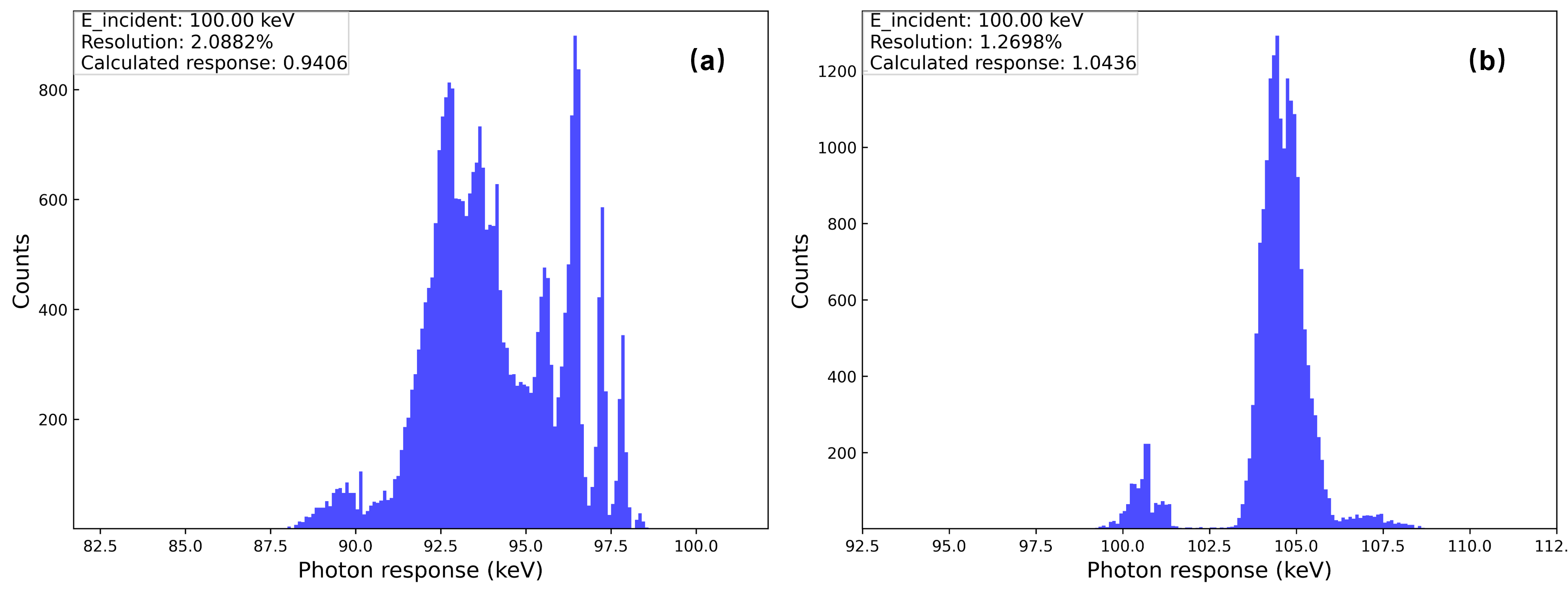}
\caption{Reconstructed energy response distributions of the LaBr$_3$(Ce,Sr) (a) and NaI(Tl) (b) crystals for 100 keV gamma photons are presented. The figures also display the the contribution of non-proportional luminescence to the total energy resolution (Resolution), and the reconstructed relative responses (Calculated response).}
\label{fig:12}
\end{figure*}

Figure~\ref{fig:12} shows the reconstructed energy response distributions of LaBr$_3$(Ce,Sr) and NaI(Tl) crystals for full-energy deposition events of 100 keV gamma photons. We observe that the energy response distribution of the gamma photons is not strictly Gaussian but exhibits some complex structures. This is due to the diverse energies of the secondary electrons generated from the interaction of gamma rays with matter. For the LaBr$_3$(Ce,Sr) crystal (Fig.~\ref{fig:12} (a)), the presence of two absorption edges and increased non-proportionality near these edges leads to more detailed structures in the energy response distribution, resembling an overlay of multiple Gaussian distributions. In the case of the NaI(Tl) crystal (Fig.~\ref{fig:12} (b)), when the gamma photon energy exceeds the binding energy of I's K-shell electrons, the response distribution exhibits two Gaussian-like peaks. However, each "Gaussian" contains finer structures. This occurs because the light yield of the NaI(Tl) crystal decreases slightly near the absorption edge. The relative photon response ($R_\gamma$, also known as Calculated response) can be computed using Equation~\ref{eq:5}. The dispersion of this distribution, characterized by the standard deviation, can be used to evaluate the contribution of non-proportionality to the total energy resolution ($\delta_{non}$) according to Equation~\ref{eq:7}. All of the results are presented in Fig.~\ref{fig:12}. The "defect" luminescence of LaBr$_3$(Ce,Sr) crystal and the "excess" luminescence of NaI(Tl) crystal are both clearly reflected in the energy response distributions \cite{FPY}, with differences in luminescence properties being one reason for the variations between the two. Another important reason for the discrepancies in energy response distributions is the notable difference in secondary electron energy distributions.
% 图8展示了重建的NaI(Tl)和LaBr3(Ce,Sr)晶体对100keV伽马光子的全能量沉积事件的能量响应分布。我们观察到伽马光子的能量响应分布不是严格的高斯型分布，而是具有一些复杂结构，这是因为伽马射线与物质相互作用后产生的次级电子能量具有多样性。对于NaI(Tl)晶体 (Fig.~\ref{fig:12}(a))，当伽马光子能量大于碘的K壳层电子结合能（33.17keV）时，该响应分布上出现了两个形似“高斯”的峰，但实际上每个“高斯”有着更精细的结构。这是因为在吸收边附近时，NaI(Tl)晶体光产额会稍微下降。由公式5可以计算出相对光子响应（R_gamma,也称为Calculated response）。该分布的离散程度用标准差σ表示，可用于衡量非线性对总能量分辨率的贡献（delta_non，公式7），结果也都展示在图8中。对于LaBr3(Ce,Sr)晶体 (Fig.~\ref{fig:12}(b))，它存在两个吸收边，并且吸收边附近伽马光子非线性增大，这导致能量响应分布中出现了较多细节结构，类似于多个高斯分布的叠加效应。LaBr3(Ce,Sr)晶体的“defect”发光与NaI(Tl)晶体“excess”发光在能量响应分布上都有明显的表现\cite{FPY}，发光特性的区别是导致二者能量响应分布差异的原因之一。次级电子能量分布的显著差异是造成能量响应分布差异的另一个重要原因。

The reconstructed energy response distribution of the LaBr$_3$(Ce,Sr) crystal shows photon response less than 1, indicative of its "defect" luminescence characteristics. This aligns well with published experimental results \cite{FPY}. For 100 keV gamma photons, the contribution of luminescence non-proportionality ($\delta_{non}$) in the LaBr$_3$(Ce,Sr) crystal is 2.09\% ± 0.00\%. The intrinsic resolution of the LaBr$_3$(Ce,Sr) crystal ($\delta_{int}$), calculated using Equation~\ref{eq:1}, is 2.45\% ± 0.05\%. The contribution of fluctuations during energy transfer ($\delta_{trans}$) is then determined to be 1.28\% ± 0.10\% using Equation~\ref{eq:2}. In contrast, the reconstructed energy response distribution of the NaI(Tl) crystal shows calculated photon response greater than 1, attributed to its "excess" luminescence. The experimentally measured non-proportional curve for X-ray responses in the 10-100 keV range also exhibits significant exceedance \cite{FPY}. The simulation results in this study are consistent with previously published experimental findings. For 100 keV gamma photons, the contribution of luminescence non-proportionality ($\delta_{non}$) in the NaI(Tl) crystal to its energy resolution is 1.27\% ± 0.00\%. The intrinsic resolution ($\delta_{int}$) and the contribution of fluctuations during energy transfer ($\delta_{trans}$), calculated using Equations~\ref{eq:1} and~\ref{eq:2}, are 2.90\% ± 0.21\% and 2.61\% ± 0.23\%, respectively.
% 对于重建的NaI(Tl)晶体的能量响应分布，呈现出大于“1”的计算响应，这是由于它的“excess”发光造成的。实验上测得的X射线非线性曲线在10-100keV能区内的相对响应都存在明显的超出。本研究中的模拟结果和已发表的实验结果具有一致性。对于100keV的伽马光子，NaI(Tl)晶体发光非线性对其能量分辨率的贡献（delta_non）为1.27%+-0.00%。根据公式1计算出能量传递过程中涨落的贡献（delta_trans）为2.61%+-0.24%。再根据公式2计算出NaI(Tl)晶体的本征分辨率（delta_int）为2.90%+-0.21%。然而，对于重建的LaBr3(Ce,Sr)晶体的能量响应分布，都呈现出小于“1”的响应，这便是它的“defect”发光特性。这与已发布的实验测试结果能较好的符合。对于100keV的伽马光子，LaBr3(Ce,Sr)晶体发光非线性对其能量分辨率的贡献（delta_non）为2.09%+-0.00%，根据公式1和2计算出能量传递过程中涨落的贡献（delta_trans）和本征分辨率（delta_int）分别是0.22+-0.12%和2.10+-0.06%。

Table~\ref{tab:3} presents the quantified results of the seven factors influencing energy resolution discussed in this study. The results indicate that the contribution of photoelectron statistical fluctuations is the largest in the energy resolutions of both LaBr$_3$(Ce,Sr) and NaI(Tl) crystals, making it the primary factor affecting energy resolution. The higher light yield of the LaBr$_3$(Ce,Sr) crystal contributes to its superior energy resolution. Another significant component is the intrinsic resolution, which ranks just below the contribution of photoelectron statistical fluctuations in both crystals. We believe the sources of intrinsic resolution are twofold: fluctuations during energy transfer and non-proportional luminescence. The results reveal a notable difference in the sources of intrinsic resolution between the LaBr$_3$(Ce,Sr) and NaI(Tl) crystals. For the LaBr$_3$(Ce,Sr) crystal, the intrinsic resolution mainly arises from non-proportional luminescence, attributable to the greater non-proportionality of the LaBr$_3$(Ce,Sr) crystal compared to NaI(Tl) crystal \cite{FPY}; however, for the NaI(Tl) crystal, it predominantly stems from fluctuations during energy transfer. This study demonstrates that the contribution of intrinsic resolution components differs significantly for crystals with varying luminescent properties.
% 表3展示了本研究所讨论的七项影响能量分辨率的因素的被量化的结果。结果表明，NaI(Tl)和 LaBr3(Ce,Sr)晶体的能量分辨率中，光电子统计涨落的贡献都占最大比重，它是影响能量分辨率最主要的原因。可以说，LaBr3(Ce,Sr)晶体较高的光产额使其具有较好的能量分辨率。另一个显著成分是本征分辨率，它在两种晶体能量分辨率中的比重仅次于光电子统计涨落的贡献。我们认为本征分辨率的来源有两个，分别是能量传递过程中的涨落和非线性的发光。结果显示，NaI(Tl)和 LaBr3(Ce,Sr)晶体本征分辨率的来源存在明显差别。对于NaI(Tl)晶体，本征分辨率主要来源于能量传递过程中的涨落；然而对于LaBr3(Ce,Sr)晶体，本征分辨率主要来源于非线性的发光，这是因为LaBr3(Ce,Sr)晶体有着比NaI(Tl)晶体更大的非线性\cite{FPY}。本研究表明，对于具有不同发光特性的晶体，其本征分辨率中的成分占比有着明显差别。

\begin{table*}
\caption{Contributions of individual components to the total energy resolution of the LaBr$_3$(Ce,Sr) and NaI(Tl) crystals at 100 keV.}
\label{tab:3}
\setlength{\tabcolsep}{3pt}
\begin{tabular}{|p{70pt}|p{230pt}|p{80pt}|p{80pt}|}
\hline
Component & Method & LaBr$_3$(Ce,Sr) & NaI(Tl) \\
\hline
$\sigma/E$ & $\sigma/E$ & 3.71\% ± 0.03\% & 4.41\% ± 0.14\% \\
$\delta_{un,radial}$ & $\sigma/\overline{x}$ & 0.11\% ± 0.01\%  & 0.12\% ± 0.01\% \\
$\delta_{un,vertical}$ & $\sigma/\overline{x}$ & 0.07\% ± 0.01\% & 0.08\% ± 0.01\% \\
$\delta_{st}$ & $1/\sqrt{N_{photoelectron}}$ & 2.69\% ± 0.00\% & 3.20\% ± 0.00\% \\
$\delta_{spe}$ & $(\sigma/E)_{spe}/\sqrt{N_{photoelectron}}$ & 0.73\% ± 0.05\% & 0.87\% ± 0.06\% \\
$\delta_{noise}$ & High-performance PMT. & Disregard & Disregard \\
$\delta_{temp}$ & Ambient temperature was maintained at 20±1°C. & Disregard & Disregard \\

$\delta_{int}$ & $\sqrt{(\sigma/E)^2-\delta_{un,radial}^2-\delta_{un,vertical}^2-\delta_{st}^2-\delta_{spe}^2}$ & 2.45\% ± 0.05\% & 2.90\% ± 0.21\% \\

$\delta_{non}$ & $\sigma/\overline{L_\gamma}$ & 2.09\% ± 0.00\% & 1.27\% ± 0.00\% \\
$\delta_{trans}$ & $\sqrt{\delta_{int}^2-\delta_{non}^2}$ & 1.28\% ± 0.10\% & 2.61\% ± 0.23\% \\
\hline
\end{tabular}
\end{table*}

\section{Conclusion}

GECAM is a specialized gamma-ray monitor that utilizes a large number of LaBr$_3$ and NaI(Tl) crystals produced by the Beijing Glass Research Institute as sensitive materials for gamma-ray detection. In this study, we conducted a comprehensive investigation of the energy resolution of 1-inch LaBr$_3$(Ce,Sr) and NaI(Tl) crystal samples from the same production batch. We first analyzed seven factors that may influence energy resolution, then employed two Hard X-ray Calibration Facilities (HXCFs) to test the energy-channel (E-C) relationship, total energy resolution, and radial non-uniformity of the crystals. Additionally, we used a PMT single-photoelectron calibration system to quantify the contributions of photoelectron statistical fluctuations and PMT single-photoelectron resolution, and we simulated the particle transport processes using Geant4 to calculate the energy resolution caused by luminescence non-proportionality and vertical non-uniformity. The results indicate that the contributions of various components to energy resolution differ, with photoelectron statistical fluctuations and intrinsic resolution being predominant. We focused on precise and detailed experimental measurements, finding that for 100 keV X-rays, the intrinsic resolutions of the LaBr$_3$(Ce,Sr) and NaI(Tl) crystals are 2.45\% ± 0.05\% and 2.90\% ± 0.21\%, respectively. The sources of intrinsic resolution for these two crystals, which have different luminescent properties, show significant differences. The intrinsic resolution of the LaBr$_3$(Ce,Sr) crystal primarily arises from luminescence non-proportionality, while that of the NaI(Tl) crystal mainly stems from fluctuations during energy transfer. Subtracting all considered factors from the measured total energy resolution is a conservative approach, as the correlations between the components have not yet been investigated. In the future, we will focus on these more in-depth studies. This work not only addresses the fitting issues of the energy resolution curves in the ground calibration of the GECAM gamma-ray detectors but also provides insights into the ultimate resolution—intrinsic resolution—of LaBr$_3$(Ce,Sr) crystals as a novel type of detection material. This innovative research offers valuable references for manufacturers to deepen their understanding of the energy resolution of LaBr$_3$(Ce,Sr) crystals, which may aid in optimizing and improving crystal growth processes for better energy resolution.
% GECAM是一种专用的伽马射线监测器，它采用了大批量的产自北玻院的LaBr3和NaI晶体作为伽马探测器的灵敏材料。在本研究中，我们对同一批次生产的1英寸LaBr3(Ce,Sr)和NaI(Tl)晶体样品的能量分辨率进行了全面的研究。我们首先分析了可能引入能量分辨率的七种因素，然后使用硬X射线地面标定装置（HXCF）测试了晶体的能量-通道（E-C）关系、总能量分辨率和水平不均匀性，使用PMT单光电子刻度装置测试了单光电子响应以量化光电子统计涨落和PMT单光电子分辨率的贡献，并且还使用Geant4对粒子输运过程进行了模拟以计算晶体发光非线性和垂直不均匀性导致的能量分辨率。结果表明，能量分辨率中各成分的占比不一，以本征分辨率和光电子统计涨落的贡献为主。我们侧重精确而具体的实验测量，对于100keV X射线，LaBr3(Ce,Sr)和NaI(Tl)晶体的本征分辨率分别是2.10%+-0.06%和2.90%+-0.21%。这两种具有不同发光特性的晶体，它们的本征分辨率的来源有着明显的区别。LaBr3(Ce,Sr)晶体的本征分辨率主要来源于发光非线性，而NaI(Tl)晶体主要来源于能量传递过程的涨落。从测量的总能量分辨率中扣除所有被考虑到的因素是一种保守的方法，因为各成分之间的相关性还未被研究，在未来我们将致力于这些更深入的研究。本工作不仅解决了GECAM伽马探测器地面标定中能量分辨率曲线的拟合问题，也提供了LaBr3(Ce,Sr)晶体作为一种新型探测材料所具有的极限分辨率——本征分辨率。这是一项创新性研究，厂家可以参考我们的研究结果，以加深对LaBr3(Ce,Sr)晶体能量分辨率的理解，这可能有助于优化和改进晶体生长工艺以获得更好的能量分辨率。

% \section*{Acknowledgment}

% \section*{References}

\bibliography{mybibfile}

@article{deng2022exploring,
  title={Exploring the intrinsic energy resolution of liquid scintillator to approximately 1 MeV electrons},
  author={Deng, Y and Sun, X and Qi, B and others},
  journal={J. Instrum.},
  volume={17},
  number={04},
  pages={P04018},
  year={2022},
  publisher={IOP Publishing},
  note={doi: {\color{blue}\href{https://doi.org/10.1088/1748-0221/17/04/P04018}{https://doi.org/10.1088/1748-0221/17/04/P04018}}}
}

@article{kaintura2021energy,
  title={Energy resolution of Compton electrons in LaCl 3: Ce using compact digitizer},
  author={Kaintura, Sanjeet S. and Ranga, V. and Panwar, S. and others},
  journal={J. Radioanal. Nucl. Chem.},
  volume={330},
  pages={1527--1531},
  year={2021},
  publisher={Springer},
  note={doi: {\color{blue}\href{https://doi.org/10.1007/s10967-021-07942-2}{https://doi.org/10.1007/s10967-021-07942-2}}}
}

@ARTICLE{FPY,
  author={P. Y. Feng and X. L. Sun and Z. H. An and others},
  title={The Energy Response of LaBr3(Ce), LaBr3(Ce,Sr), and NaI(Tl) Crystals for GECAM},
  journal={Nucl. Sci. Tech.},
  note={doi: {\color{blue}\href{https://doi.org/10.1007/s41365-024-01383-8}{https://doi.org/10.1007/s41365-024-01383-8}}}
}

@article{LI2014Particle,
  title={Particle Discrimination Measurement of Liquid Scintillators Using DT5751},
  author={Li, X. and Ren, J. and Ruan, X.C. and others},
  journal={Ann. Rep. Chin. Inst. Atom. En.},
  volume={00},
  pages={147-147},
  year={2014},
  note={doi: {\color{blue}\href{https://doi.org/CNKI:SUN:YNXB.0.2014-00-055}{https://doi.org/CNKI:SUN:YNXB.0.2014-00-055}}}
}

@article{LIMKITJAROENPORN201815110,
title = {The light yield non-proportionality and electron energy resolution study of CsI(Tl) scintillator by Compton coincidence technique (CCT)},
journal = {Mat. Today-Proc.},
volume = {5},
number = {7, Part 1},
pages = {15110-15114},
year = {2018},
issn = {2214-7853},
doi = {https://doi.org/10.1016/j.matpr.2018.04.066},
url = {https://www.sciencedirect.com/science/article/pii/S2214785318307442},
author = {P. Limkitjaroenporn and W. Hongtong and W. Chaiphaksa and others},
note={doi: {\color{blue}\href{https://doi.org/10.1016/j.matpr.2018.04.066}{https://doi.org/10.1016/j.matpr.2018.04.066}}}
}

@article{wei2018consistency,
  title={Consistency test of PMT SPE spectrum from dark-noise pulses and LED low-intensity light},
  author={ Wei, Y.T. and Guan, M.Y. and Xiong, W.X. and others},
  journal={Radiat. Detect. Technol. Methods},
  volume={2},
  number={1},
  pages={11},
  year={2018},
  note={doi: {\color{blue}\href{https://doi.org/10.1007/s41605-018-0042-6}{https://doi.org/10.1007/s41605-018-0042-6}}}
}

@article{2021Ground,
  title={Ground-based calibration and characterization of LaBr3-SiPM-based gamma-ray detector on GECAM satellite: 8--160 keV},
  author={He, J.J. and An, Z.H. and Peng, W.X. and others},
  journal={Mon. Not. R. Astron. Soc.},
  volume={525},
  number={3},
  pages={3399--3412},
  year={2023},
  publisher={Oxford University Press},
  note={doi: {\color{blue}\href{https://doi.org/10.1093/mnras/stad2439}{https://doi.org/10.1093/mnras/stad2439}}}
}

@article{2021Calibration,
  title={Calibration study of the Gamma-Ray Monitor onboard the SVOM satellite},
  author={ Wen, X. and  Sun, J.C. and  He, J. and others},
  journal={Nucl. Instrum. Methods Phys. Res. Sect. A-Accel. Spectrom. Dect. Assoc. Equip.},
  number={1},
  pages={165301},
  year={2021},
  note={doi: {\color{blue}\href{https://doi.org/10.1016/j.nima.2021.165301}{https://doi.org/10.1016/j.nima.2021.165301}}}
}

@article{2019Ground,
  title={Ground-based calibration and characterization of the HE detectors for Insight-HXMT},
  author={ Li, X.F. and Liu, C.Z. and Chang, Z. and others},
  journal={J. High. Energy Astrophys.},
  volume={24},
  year={2019},
  note={doi: {\color{blue}\href{https://doi.org/10.1016/j.jheap.2019.09.003}{https://doi.org/10.1016/j.jheap.2019.09.003}}}
}

@article{2016LEGe,
  title={LEGe detector intrinsic efficiency calibration for parallel incident photons},
  author={ H.R., Liu  and  J.J., Wu  and  J.C., Liang  and  others},
  journal={Appl. Radiat. Isot.},
  volume={109},
  pages={551-554},
  year={2016},
  note={doi: {\color{blue}\href{https://doi.org/10.1016/j.apradiso.2015.11.1023}{https://doi.org/10.1016/j.apradiso.2015.11.102}}}
}

@article{Ground-calibration-GECAM-C,
  title={Ground calibration of gamma-ray detectors of GECAM-C},
  author={Zheng, C. and An, Z.H. and Peng, W.X. and others},
  journal={Nucl. Instrum. Methods Phys. Res. Sect. A-Accel. Spectrom. Dect. Assoc. Equip.},
  volume={1059},
  pages={169009},
  year={2024},
  publisher={Elsevier},
  note={doi: {\color{blue}\href{https://doi.org/10.1016/j.nima.2023.169009}{https://doi.org/10.1016/j.nima.2023.169009}}}
}

@article{2017Intrinsic,
  title={Intrinsic Resolution Of Compton Electrons in CeBr 3 scintillator using compact CCT},
  author={Ranga, V. and Rawat, S. and Sharma, S. and others},
  journal={IEEE Trans. Nucl. Sci.},
  volume={65},
  number={1},
  pages={616--620},
  year={2017},
  publisher={IEEE},
  note={doi: {\color{blue}\href{https://doi.org/10.1109/TNS.2017.2779888}{https://doi.org/10.1109/TNS.2017.2779888}}}
}

@inproceedings{2009A,
  title={A technique for measuring the energy resolution of low-Z scintillators},
  author={Roemer, K. and Pausch, G. and Herbach, C.M. and others},
  booktitle={IEEE NSS/MIC 2009},
  pages={6--11},
  year={2009},
  organization={IEEE},
  note={doi: {\color{blue}\href{https://doi.org/10.1109/NSSMIC.2009.5401909}{https://doi.org/10.1109/NSSMIC.2009.5401909}}}
}

@article{2012Non,
  title={Non-proportionality of electron response and energy resolution of Compton electrons in scintillators},
  author={Swiderski, L. and Marcinkowski, R. and Szawlowski, M. and others},
  journal={IEEE Trans. Nucl. Sci.},
  volume={59},
  number={1},
  pages={222--229},
  year={2012},
  publisher={IEEE},
  note={doi: {\color{blue}\href{https://doi.org/10.1109/TNS.2011.2175407}{https://doi.org/10.1109/TNS.2011.2175407}}}
}

@article{moszynski2016energy,
  title={Energy resolution of scintillation detectors},
  author={Moszyski, M. and Syntfeld-Kauch, A. and Swiderski, L. and others},
  journal={Nucl. Instrum. Methods Phys. Res. Sect. A-Accel. Spectrom. Dect. Assoc. Equip.},
  volume={805},
  pages={25--35},
  year={2016},
  publisher={Elsevier},
  note={doi: {\color{blue}\href{https://doi.org/10.1016/j.nima.2015.07.059}{https://doi.org/10.1016/j.nima.2015.07.059}}}
}

@article{han2023csi,
  title={CsI-bowl: an ancillary detector for exit channel selection in $\gamma$-ray spectroscopy experiments},
  author={Han, X. C. and Wang, S. and Wu, H. Y. and others},
  journal={Nucl. Sci. Tech.},
  volume={34},
  number={9},
  pages={133},
  year={2023},
  publisher={Springer},
  note={doi: {\color{blue}\href{https://doi.org/10.1007/s41365-023-01289-x}{https://doi.org/10.1007/s41365-023-01289-x}}}
}

@article{liu2023toward,
  title={Toward real-time digital pulse process algorithms for CsI (Tl) detector array at external target facility in HIRFL-CSR},
  author={Liu, T. and Song, H.S. and Yu, Y.H. and others},
  journal={Nucl. Sci. Tech.},
  volume={34},
  number={9},
  pages={131},
  year={2023},
  publisher={Springer},
  note={doi: {\color{blue}\href{https://doi.org/10.1007/s41365-023-01272-6}{https://doi.org/10.1007/s41365-023-01272-6}}}
}

@article{zhang2022transition,
  title={Transition edge sensor-based detector: from X-ray to $\gamma$-ray},
  author={Zhang, S. and Xia, J. K. and Sun, T. and others},
  journal={Nucl. Sci. Tech.},
  volume={33},
  number={7},
  pages={84},
  year={2022},
  publisher={Springer},
  note={doi: {\color{blue}\href{https://doi.org/10.1007/s41365-022-01071-5}{https://doi.org/10.1007/s41365-022-01071-5}}}
}

@article{lu2022monte,
  title={Monte Carlo simulation for performance evaluation of detector model with a monolithic LaBr3 (Ce) crystal and SiPM array for $\gamma$ radiation imaging},
  author={Lu, W. and Wang, L. and Yuan, Y. and others},
  journal={Nucl. Sci. Tech.},
  volume={33},
  number={8},
  pages={107},
  year={2022},
  publisher={Springer},
  note={doi: {\color{blue}\href{https://doi.org/10.1007/s41365-022-01081-3}{https://doi.org/10.1007/s41365-022-01081-3}}}
}

@article{SSPMA-2019-0417,
  author = "X. Q. Li and X. Y. Wen and Z. H. An and others",
  title = "The GECAM and its payload",
  journal = "Sci. Sin.-Phys. Mech. Astron.",
  year = "2020",
  volume = "50",
  number = "12",
  pages = "129508-",
  doi = "https://doi.org/10.1360/SSPMA-2019-0417",
  note={doi: {\color{blue}\href{https://doi.org/10.1360/SSPMA-2019-0417}{https://doi.org/10.1360/SSPMA-2019-0417}}}
}

@article{SSPMA-2020-0457,
  author = "S. L. Xiong",
  title = "Special Topic: GECAM gamma-ray all-sky monitor",
  journal = "Sci. Sin.-Phys. Mech. Astron.",
  year = "2020",
  volume = "50",
  number = "12",
  pages = "129501-",
  doi = "https://doi.org/10.1360/SSPMA-2020-0457",
  note={doi: {\color{blue}\href{https://doi.org/10.1360/SSPMA-2020-0457}{https://doi.org/10.1360/SSPMA-2020-0457}}}
}

@article{an2023insight,
  title={Insight-HXMT and GECAM-C observations of the brightest-of-all-time GRB 221009A},
  author={Z. H. An and S. Antier and X. Z. Bi and others},
  journal={arXiv preprint arXiv:2303.01203},
  year={2023},
 doi = "https://doi.org/10.48550/arXiv.2303.01203",
  note={doi: {\color{blue}\href{https://doi.org/10.48550/arXiv.2303.01203}{https://doi.org/10.48550/arXiv.2303.01203}}}
}

@article{sun2023magnetar,
  title={Magnetar emergence in a peculiar gamma-ray burst from a compact star merger},
  author={H. Sun and C. W. Wang and J. Yang and others},
  journal={arXiv preprint arXiv:2307.05689},
  year={2023},
  doi = "https://doi.org/10.48550/arXiv.2307.05689",
  note={doi: {\color{blue}\href{https://doi.org/10.48550/arXiv.2307.05689}{https://doi.org/10.48550/arXiv.2307.05689}}}
}

@article{zhang2022dedicated,
  title={Dedicated SiPM array for GRD of GECAM},
  author={D. L. Zhang and X. L. Sun and Z. H. An and others},
  journal={Radiat. Detect. Technol. Methods},
  volume={6},
  number={1},
  pages={63--69},
  year={2022},
  publisher={Springer},
  doi = "https://doi.org/10.1007/s41605-021-00299-w",
  note={doi: {\color{blue}\href{https://doi.org/10.1007/s41605-021-00299-w}{https://doi.org/10.1007/s41605-021-00299-w}}}
}

@article{zhang2019energy,
  title={Energy response of GECAM gamma-ray detector based on LaBr3: Ce and SiPM array},
  author={D. L. Zhang and X. Q. Li and S. L. Xiong and others},
  journal={Nucl. Instrum. Methods Phys. Res. Sect. A-Accel. Spectrom. Dect. Assoc. Equip.},
  volume={921},
  pages={8--13},
  year={2019},
  publisher={Elsevier},
  doi = "https://doi.org/10.1016/j.nima.2018.12.032",
  note={doi: {\color{blue}\href{https://doi.org/10.1016/j.nima.2018.12.032}{https://doi.org/10.1016/j.nima.2018.12.032}}}
}

@article{zhang2023performance,
  title={The performance of SiPM-based gamma-ray detector (GRD) of GECAM-C},
  author={D. L. Zhang and C. Zheng and J. C. Liu and others},
  journal={Nucl. Instrum. Methods Phys. Res. Sect. A-Accel. Spectrom. Dect. Assoc. Equip.},
  volume={1056},
  pages={168586},
  year={2023},
  publisher={Elsevier},
  doi = "https://doi.org/10.1016/j.nima.2023.168586",
  note={doi: {\color{blue}\href{https://doi.org/10.1016/j.nima.2023.168586}{https://doi.org/10.1016/j.nima.2023.168586}}}
}

@article{feng2024detector,
  title={Detector performance of the Gamma-ray Transient Monitor onboard DRO-A Satellite},
  author={Feng, P. Y. and An, Z. H. and Zhang, D. L. and others},
  journal={Sci. China-Phys. Mech. Astron.},
  volume={67},
  number={11},
  pages={1--16},
  year={2024},
  publisher={Springer},
  doi = "https://doi.org/10.1007/s11433-024-2458-9",
  note={doi: {\color{blue}\href{https://doi.org/10.1007/s11433-024-2458-9}{https://doi.org/10.1007/s11433-024-2458-9}}}
}

@article{wang2024simulation,
  title={Simulation of the in-flight background and performance of DRO/GTM},
  author={C. W. Wang and J. Zhang and S. J. Zheng and others},
  journal={Exp. Astron.},
  volume={57},
  number={3},
  pages={26},
  year={2024},
  publisher={Springer},
  doi = "https://doi.org/10.1007/s10686-024-09946-8",
  note={doi: {\color{blue}\href{https://doi.org/10.1007/s10686-024-09946-8}{https://doi.org/10.1007/s10686-024-09946-8}}}
}

@article{zheng2024observation,
  title={Observation of GRB 221009A Early Afterglow in X-Ray/Gamma-Ray Energy Bands},
  author={C. Zheng and Y. Q. Zhang and S. L. Xiong and others},
  journal={Astrophys. J. Lett.},
  volume={962},
  number={1},
  pages={L2},
  year={2024},
  publisher={IOP Publishing},
  doi = "https://doi.org/10.3847/2041-8213/ad2073",
  note={doi: {\color{blue}\href{https://doi.org/10.3847/2041-8213/ad2073}{https://doi.org/10.3847/2041-8213/ad2073}}}
}

@article{li2021technology,
  title={The technology for detection of gamma-ray burst with GECAM satellite},
  author={X. Q. Li and X. Y. Wen and Z. H. An and others},
  journal={Radiat. Detect. Technol. Methods},
  pages={1--14},
  year={2021},
  publisher={Springer},
  doi = "https://doi.org/10.1007/s41605-021-00288-z",
  note={doi: {\color{blue}\href{https://doi.org/10.1007/s41605-021-00288-z}{https://doi.org/10.1007/s41605-021-00288-z}}}
}

@article{zhang2020overview,
  title={Overview to the hard X-ray modulation telescope (Insight-HXMT) satellite},
  author={S. N. Zhang and T. P. Li and F. J. Lu and others},
  journal={Sci. China-Phys. Mech. Astron.},
  volume={63},
  pages={1--18},
  year={2020},
  publisher={Springer},
  doi = "https://doi.org/10.1007/s11433-019-1432-6",
  note={doi: {\color{blue}\href{https://doi.org/10.1007/s11433-019-1432-6}{https://doi.org/10.1007/s11433-019-1432-6}}}
}

@article{huyan2018geant4,
  title={Geant4 simulations of the absorption of photons in CsI and NaI produced by electrons with energies up to 4 MeV and their application to precision measurements of the $\beta$-energy spectrum with a calorimetric technique},
  author={X. Huyan and O. Naviliat-Cuncic and P. Voytas and others},
  journal={Nucl. Instrum. Methods Phys. Res. Sect. A-Accel. Spectrom. Dect. Assoc. Equip.},
  volume={879},
  pages={134--140},
  year={2018},
  publisher={Elsevier},
  doi = "https://doi.org/10.1016/j.nima.2017.10.061",
  note={doi: {\color{blue}\href{https://doi.org/10.1016/j.nima.2017.10.061}{https://doi.org/10.1016/j.nima.2017.10.061}}}
}

@article{payne2015nonproportionality,
  title={Nonproportionality of scintillator detectors. IV. Resolution contribution from delta-rays},
  author={S. A. Payne},
  journal={IEEE Trans. Nucl. Sci.},
  volume={62},
  number={1},
  pages={372--380},
  year={2015},
  publisher={IEEE},
  doi = "https://doi.org/10.1109/TNS.2014.2387256",
  note={doi: {\color{blue}\href{https://doi.org/10.1109/TNS.2014.2387256}{https://doi.org/10.1109/TNS.2014.2387256}}}
}

@article{payne2011nonproportionality,
  title={Nonproportionality of scintillator detectors: Theory and experiment. II},
  author={S. A. Payne and W. W. Moses and S. Sheets and others},
  journal={IEEE Trans. Nucl. Sci.},
  volume={58},
  number={6},
  pages={3392--3402},
  year={2011},
  publisher={IEEE},
  doi = "https://doi.org/10.1109/TNS.2011.2167687",
  note={doi: {\color{blue}\href{https://doi.org/10.1109/TNS.2011.2167687}{https://doi.org/10.1109/TNS.2011.2167687}}}
}

@article{sriwongsa2019non,
  title={Non-Proportionality Electron Response and Energy Resolution of LaBr 3: Ce and LuYAP: Ce Scintillating Crystals},
  author={K. Sriwongsa and P. Limkitjaroenporn and W. Hongtong and others},
  journal={J. Korean Phys. Soc.},
  volume={75},
  pages={672--677},
  year={2019},
  publisher={Springer},
  doi = "https://doi.org/10.3938/jkps.75.672",
  note={doi: {\color{blue}\href{https://doi.org/10.3938/jkps.75.672}{https://doi.org/10.3938/jkps.75.672}}}
}

@article{swiderski2010energy,
  title={Energy resolution of Compton electrons in LaBr $ \_ $\{$3$\}$ $: Ce scintillator},
  author={L. Swiderski and M. Moszynski and W. Czarnacki and others},
  journal={IEEE Trans. Nucl. Sci.},
  volume={57},
  number={3},
  pages={1697--1701},
  year={2010},
  publisher={IEEE},
  doi = "https://doi.org/10.1109/TNS.2010.2045899",
  note={doi: {\color{blue}\href{https://doi.org/10.1109/TNS.2010.2045899}{https://doi.org/10.1109/TNS.2010.2045899}}}
}

@article{moszynski2004intrinsic,
  title={Intrinsic energy resolution and light yield nonproportionality of BGO},
  author={M. Moszynski and M. Balcerzyk and W. Czarnacki and others},
  journal={IEEE Trans. Nucl. Sci.},
  volume={51},
  number={3},
  pages={1074--1079},
  year={2004},
  publisher={IEEE},
  note={doi: {\color{blue}\href{https://doi.org/10.1109/TNS.2004.829491}{https://doi.org/10.1109/TNS.2004.829491}}},
  doi={10.1109/TNS.2004.829491}
}

@article{bissaldi2009ground,
  title={Ground-based calibration and characterization of the Fermi gamma-ray burst monitor detectors},
  author={E. Bissaldi and A. von Kienlin and G. Lichti and others},
  journal={Exp. Astron.},
  volume={24},
  pages={47--88},
  year={2009},
  publisher={Springer},
  note={doi: {\color{blue}\href{https://doi.org/10.1007/s10686-008-9135-4}{https://doi.org/10.1007/s10686-008-9135-4}}}
}

@article{2022Electron,
  title={Electron non-linear light yield of LaBr3 detector aboard GECAM},
  author={C. Zheng and W. X. Peng and X. B. Li and  others},
  journal={Nucl. Instrum. Methods Phys. Res. Sect. A-Accel. Spectrom. Dect. Assoc. Equip.},
  year={2022},
  note={doi: {\color{blue}\href{https://doi.org/10.1016/j.nima.2022.167427}{https://doi.org/10.1016/j.nima.2022.167427}}}
}

@article{valentine1998light,
  title={The light yield nonproportionality component of scintillator energy resolution},
  author={J. D. Valentine and B. D. Rooney and J. Li},
  journal={IEEE Trans. Nucl. Sci.},
  volume={45},
  number={3},
  pages={512--517},
  year={1998},
  publisher={IEEE},
  note={doi: {\color{blue}\href{https://doi.org/10.1109/23.682438}{https://doi.org/10.1109/23.682438}}}
}

@article{zhao2023gecam,
  title={Gecam localization of high-energy transients and the systematic error},
  author={Y. Zhao and W. C. Xue and S. L. Xiong and others},
  journal={Astrophys. J. Suppl. Ser.},
  volume={265},
  number={1},
  pages={17},
  year={2023},
  publisher={IOP Publishing},
  note={doi: {\color{blue}\href{https://doi.org/10.3847/1538-4365/acafeb}{https://doi.org/10.3847/1538-4365/acafeb}}},
  doi={10.3847/1538-4365/acafeb}
}

@article{zhao2023paired,
  title={Paired quasi-periodic pulsations of hard X-ray emission in a solar flare},
  author={H. S. Zhao and D. Li and S. L. Xiong and others},
  journal={Sci. China-Phys. Mech. Astron.},
  volume={66},
  number={5},
  pages={259611},
  year={2023},
  publisher={Springer},
  note={doi: {\color{blue}\href{https://doi.org/10.1007/s11433-022-2064-6}{https://doi.org/10.1007/s11433-022-2064-6}}},
  doi={10.1007/s11433-022-2064-6}
}

@article{feng2023intrinsic,
  title={The Intrinsic Energy Resolution of LaBr $ \_3 $(Ce) Crystal for GECAM},
  author={P. Y. Feng and X. L. Sun and C. E. Wang and others},
  journal={arXiv preprint arXiv:2401.00226},
  year={2023},
  note={doi: {\color{blue}\href{
https://doi.org/10.48550/arXiv.2401.00226}{
https://doi.org/10.48550/arXiv.2401.00226}}}
}

\end{document}